\documentclass[useAMS,usenatbib,referee]{mn2e}
\usepackage{graphicx}
\usepackage{epstopdf}
\usepackage[figuresright]{rotating}
\usepackage{subfigure,color}
\usepackage{longtable}
\usepackage{tabularx,multicol,lipsum}
\usepackage{xspace}

\newcommand{\ion}[2]{{\textrm{#1}}{\textrm{\sc #2}}}

\definecolor{pink}{rgb}{.9,.2,.5}  
\definecolor{purple}{rgb}{.5,.6,.7}

\title[Lyman-$\alpha$ emitter objects at high redshift]
{Nature and chemical abundances of a sample of Lyman-$\alpha$ emitter objects at high redshift
}

\author[Dors et al.]
            {O.~L.\ Dors$^{1}$\thanks{E-mail:olidors@univap.br}, 
	       B. Agarwal$^{2}$, G.~F.\ H\"agele$^{3,4}$,  M.~V. Cardaci$^{3,4}$, Claes-Erik Rydberg$^{2}$,  
	         \newauthor{R.~A. Riffel$^5$, A.~S. Oliveira$^1$, A.~C. Krabbe$^1$}
               \\
$^1$ Universidade do Vale do Para\'iba, Av. Shishima Hifumi, 2911, Cep
12244-000, S\~ao Jos\'e dos Campos, SP, Brazil\\ 
$^2$ Universit\"at Heidelberg, Zentrum f\"ur Astronomie, Institut f\"ur Theoretische Astrophysik, Albert-Ueberle-Str. 2, D-69120 Heidelberg \\
$^3$ Instituto de Astrof\'isica de La Plata (CONICET-UNLP), Argentina. \\
$^4$ Facultad de Ciencias Astron\'omicas y Geof\'{\i}sicas, Universidad Nacional de La Plata, Paseo del Bosque s/n, 1900 La Plata, Argentina.\\
$^5$ Universidade Federal de Santa Maria, Av. Roraima 1000, Cep 97105-900, Santa Maria, Brazil} 
\begin{document}

\date{Accepted 2015 Month  00. Received 2015 Month 00; in original form 2014 December 17}

\pagerange{\pageref{firstpage}--\pageref{lastpage}} \pubyear{2011}

\maketitle

\label{firstpage}

\begin{abstract}
{We built  a grid of photoionization models and compiled already available observational emission line intensities  ($\rm 1000 \: < \: \lambda(\AA) \: < \: 2000$)
of confirmed star formation regions and Active Galactic Nucleus (AGNs)  in order to classify  five Ly$\alpha$ emitter (LAE) objects at high  redshift
$(5.7 \: < \: z \:  < \:7.2)$.} We selected objects for which at least one metal emission-line was measured.
The resulting sample  is composed by the objects 
RXC\,J2248.7-4431-ID3, HSC\,J233408+004403, COSY, A1703-zd6, and CR7 (clump C). The photoionization models were built assuming a Power Law (associated with the presence of an  AGN), 
a Direct Collapse Black Hole (DCBH), and Population II stars for the ionizing source. The resulting models were then compared with observational emission-line ratios in six diagnostic 
diagrams to produce a spectral classification of the sample.
We found that CR7 (clump C), HSC\,J233408+004403 and  COSY probably have a non thermal
ionizing source (AGN or DCBH) while the RXC J2248.7-4431-ID3  and A1703-zd6
 seem to host a stellar cluster. Detailed photoionization models were constructed to reproduce 
observational emission line ratios  of the sample of LAEs, and
to derive chemical abundances and number of ionizing photons $Q(\rm H)$ of these objects.
From these models, we found  metallicities in the range  $0.1 \: \la \: (Z/Z_{\odot}) \: \la \: 0.5$
and $\log Q(\rm H) \: > \: 53$.
Values for C/O abundance ratio  derived for the LAEs seem  to be consistent with those
derived for local star forming objects with similar  metallicities, while  an overabundance of N/O  was found
for most of the LAEs.  

\end{abstract}

\begin{keywords}
galaxies: active  -- galaxies:  abundances -- galaxies: evolution -- galaxies: nuclei --
galaxies: formation-- galaxies: ISM -- galaxies: Seyfert
\end{keywords}


\section{Introduction}
\label{intro}

The study of Lyman-$\alpha$  emitter objects (hereafter LAEs) at high redshift ($z \: > 5$) plays a key role in the understanding the reionization epoch of the Universe, chemical evolution and galaxy formation.
 
In their seminal work,   \citet{partridge67}  predicted the existence of newly
formed  and highly luminous  ($\sim 3\times 10^{46}$ erg/s) galaxies  present
in an epoch when the Universe was   150 million  years old (see also \citealt{pritchet94}).  Owing to the  advancement of large ground-based telescopes (e.g. SUBARU; Very Large Telescope-VLT)  
together with the large amount of data produced by the Hubble Space Telescope (HST), several observational works 
have revealed the existence of these primeval galaxies,  commonly referred to as  LAEs 
(see e.g. \citealt{ouchi08, ouchi09, ota08, zheng09, lehnert10, stark10, 
zabl15, sobral15, oesch15,  rydberg15, rydberg17,
schmidt17,  mainali17,        
shibuya18, laporte17}, among others). In particular, spectroscopic
data of LAEs have shown that these objects, in fact, exhibit a strong Ly$\alpha$ emission line [L(Ly$\alpha)$]
but show a weak emission of \ion{He}{ii} $\lambda$1640 \AA.  Additionally, some objects exhibit metal emission lines in their spectra such as \ion{N}{v}$\lambda$1239\AA, \ion{C}{iv}$\lambda$1550\AA \,  and 
\ion{O}{iii}]$\lambda$1666\AA \, (see \citealt{matthee17, shibuya18}).

Emission-line intensities and their ratios can be used to investigate the nature of the gas emission, i. e., 
to classify an object as a Star Forming region (SF) or an Active Galactic Nucleus (AGN). This methodology, 
based on optical emission lines,  was proposed by  \citet{baldwin81}\footnote{\citet{baldwin81}
considered four classes of objects: normal \ion{H}{ii} regions, planetary nebulae, AGNs, and 
objects excited by shock-wave heating.} and hinges on the idea that the gas in AGNs presents a higher excitation degree and metallicity than
the gas in SFs.  Alternative diagnostic diagrams based on 
emission-line ratios have been proposed, e.g. in the near-infrared (IR) 
\citep{neunanen02, ardila04, riffel13, colina15} and in the ultraviolet  \cite[UV;][]{allen98, feltre16, nakajima17}.
 
Studies based on  diagnostic diagrams comprising of emission lines and on photometric data  
 of LAEs have shown  that there is no consensus
on the nature of these objects. For example,  using   the 
X-SHOOTER and  SINFONI instruments mounted on the VLT, DEIMOS on the Keck telescope, the near-IR COSMOS/UltraVISTA 
survey, and the WFPC3 on board the HST, \citet{sobral15}  obtained spectroscopic and photometric data in the UV
restframe of  CR7 ($z=6.604$). Using the high spatial resolution of HST data, they showed that CR7 is composed of three different spatially resolved components, namely A, B and C. 
They detected a strong Ly$\alpha$ and \ion{He}{ii}$\lambda$1640\AA\ emission,  but their spectra did not show metal emission lines. 
Moreover, using this data together with IRAC/Spitzer mid-IR photometric data (optical data in the CR7 restframe) 
they concluded that a combination of population III stars which dominate the rest-frame UV light and the nebular emission lines,  and a more metal enriched 
stellar population that dominates the mass, is the best explanation for the composition of CR7.
However, \cite{bowler+17}, using almost the same data as \cite{sobral15} together with deeper IR and optical photometric data, arrived at a very 
different conclusion. They claimed that a low-mass narrow-line AGN or a young low metallicity stellar cluster with super solar $\alpha$-element abundance could be the ionizing source in CR7. 
On the other hand, \citet{agarwal17}, by comparing  the observed  $H_{160}$-[3.6] and [3.6]-[4.5]  IR colors of CR7 obtained by \cite{bowler+17} with photoionization model predictions, derived that the  
clump A potentially hosts an evolved direct collapse black hole (DCBH), while components B and C are consistent with metal enriched star forming regions
(see also \citealt{pallottini15, pacucci17}). 
Recently, \citet{sobral17} presented new data for CR7 and concluded that  this object  seems to be actively forming
stars,  without any clear AGN activity in clumps A and B,  with component A experiencing the most massive starburst.
Their observations show that the component C hosts a strong ionization source, possibly an AGN (see also \citealt{visbal16}). 
Another  disagreement is found, for example, for the  LAE A1703-zd6 (at $z=7.043$).
\citet{stark15} showed that photoionization models, considering both 
a non-thermal ionizing source (AGN) and stellar clusters can reproduce the observational data of this object, thus
indicating a strong degeneracy of the results. They further argue that additional constraints on high ionization emission lines (e.g. \ion{N}{v}$\lambda$1239\AA, \ion{C}{iv}$\lambda$1550)
are needed in order to clarify the LAE's precise nature.

The lack of  \ion{He}{ii} emission,  the presence of narrow emission lines (with Full Width at Half Maximum (FWHM) between
 $\sim$100 and $\sim$600 km\,s$^{-1}$) and the low Ly$\alpha$ equivalent widths ($\rm EW(Ly\alpha) \: \la \: 200 \: \AA$) for  most part of the LAEs  (see \citealt{matthee17, shibuya18, leclercq17})
have been  used as  arguments favouring the low-metallicity stellar populations as the ionizing source of these objects
(e.g. \citealt{mainali17}).
However, the inability of SF in producing considerable Lyman continuum radiation into the intergalactic medium suggests that a significant contribution of ionizing radiation may emerge from AGNs
(e.g. \citealt{madau15, pallottini15, roberts16, laporte17}). Moreover,  
 Seyfert 2 type AGNs also show narrow-line profiles ($\rm 50 \: \la \: FWHM \: \la \: 1200$, e.g \citealt{koski78, bergeron81, dopita15})
 and low  equivalent widths ($\rm EW(Ly\alpha) \: \la \: 300 \AA$, e.g. \citealt{bergeron81, thuan84, derobertis88}), 
  indicating that LAEs could host some  kind of  non-thermal ionizing source.
 In summary, the exact nature of the LAEs still eludes us and remains an open challenge.

With the above in mind, in this paper we compare observational spectroscopic 
data of five LAEs, taken from the literature,   with results of photoionization models generated by considering distinct types of ionizing sources.
Our main goals are:
\begin{enumerate}
\item Classifying  LAEs at redshift $z\: > 5$.
\item Determining the physical parameters that can best describe LAEs (e.g. metallicity,  number of ionizing photons).
\end{enumerate} 
 This paper is organized as follows. In Section~\ref{obs}, we present the
observational data used in this study and in Section~\ref{mod},
we describe our photoionization models.
In Section~\ref{res}, the resulting classification and  physical 
parameters that can help ascertain the nature of LAEs  is presented. Finally, the discussion and conclusions are outlined in Sections \ref{disc} and \ref{conc} respectively.

 \section{Observational data}
 \label{obs}
 In what follows, we describe the LAE data and the control sample. 
We compiled UV emission-line  fluxes
 ($\rm 1000 \: < \: \lambda(\AA) \: < \: 2000$) for a sample of
 LAEs located at redshift $z \: > \:5$. Also, with the goal
 of creating a control sample, UV  emission-line  fluxes of confirmed
 SFs, Seyfert 2 nuclei, Quasars and Radio Galaxies  were  also gathered from the literature.

\subsection{LAEs}
\label{dlae} 

Our LAE selection criteria was as follows:
\begin{enumerate}
\item each selected object must be at $z \: > \: 5$,
\item the flux of the Ly$\alpha$ $\lambda$1216\AA\ emission-line was measured, 
\item the flux of \ion{He}{ii} $\lambda$1640\AA\  was measured or at least an upper limit was estimated, 
\item {the flux of  at least  one metal emission line of  \ion{N}{v} $\lambda$1239\AA, \ion{C}{iv}  $\lambda$1550\AA, \ion{O}{iii}]$\lambda$1666\AA, 
\subitem or  \ion{C}{iii}]$\lambda$1909\AA\,   was measured and, at least, one upper  limit of the \subitem flux of one metal emission-line was estimated.} 
\end{enumerate}
Throughout the paper,  \ion{C}{iv} $\lambda$1550\AA\ refers to the sum of  \ion{C}{iv} $\lambda$1548.18\AA\
 and \ion{C}{iv} $\lambda$1550.77\AA; \ion{N}{v}$\lambda$1240\AA\ is the sum
 of \ion{N}{v}$\lambda$1238.8\AA\ and  \ion{N}{v}$\lambda$1242.78\AA; and
 \ion{O}{iii}] $\lambda$1666\AA\ is the sum of \ion{O}{iii}] $\lambda$1660.81\AA\ and 
 \ion{O}{iii}] $\lambda$1666.15\AA. The selection criteria
 described above were employed so as to include each object in at least one diagnostic diagram involving metal emission lines.  

The selection resulted in  five objects, namely 
 RXC\,J2248.7-4431-ID3 (hereafter ID3), CR7 (clump C), HSC\,J233408+004403 (hereafter J233408),  COSY, and A1703-zd6.  
In Table~\ref{tab1} we summarise  the  redshift,   emission-line  fluxes, and the bibliographic
 reference for each  object in our sample. In many cases, the study from which the data were extracted present only the upper limit estimates of the 
 flux for some emission lines. This procedure is adopted by the authors because,
 often, a given  line is not  detected above  $\sim 3 \: \sigma$ (e.g.\  \ion{C}{iv}$\lambda$1550\AA\, in CR7;  \citealt{sobral17}).
 In particular, for A1703-zd6, only the \ion{O}{iii}]$\lambda$1660.81\AA\ line  of the  doublet was measured by \citep{stark15}. {Thus, for this case, we  adopted the theoretical    \ion{O}{iii}]($\lambda1660.81$\AA/$\lambda1666.15$\AA)=0.34 line flux ratio  
to calculate the flux of \ion{O}{iii}] $\lambda$1666\AA.} 
 In what follows, we summarize the results obtained in previous works for 
 the considered LAEs.

 \subsubsection{ID3}

\citet{schmidt17}  used  the  Hubble Grism Lens-Amplified Survey from Space  (GLASS),
to obtain photometric and spectroscopic data of components of the system  RXC\,J2248.7-4431 ($z=6.11$),
where the \ion{C}{iv} emission was detected at 3-5 $\sigma$ level for two components. They compared the constraints of   emission line flux
ratios of the components of this galaxy  with photoionization models
and found that the observational data are better reproduced by SF  
rather than AGN models. Moreover, based on {Spectral Energy Distribution} (SED) fits, the authors
found that the SF responsible for the ionization of the components
must have a mass  of $\sim 10^{9} \: M_{\odot}$, a Star Formation Rate (SFR) of $\sim 10 \: M_{\odot}/\rm yr$ and an age younger 
than 50 Myr. By using the Folded-port
InfraRed Echellette instrument coupled in the  Magellan Baade Telescope,  
\cite{mainali17} observed one component of RXCJ2248.7-443 identified by ID3.
These authors were able to measure  Ly$\alpha$  and some UV metal emission-line fluxes, however, no \ion{He}{ii} flux was detected (see Table~\ref{tab1}).
They further concluded that the hard spectrum of this object
is characteristic of low-metallicity stellar populations  and is less
consistent with AGN excitation.

 \subsubsection{CR7}
 
CR7 is one of the most luminous LAEs and 
was discovered by \citet{sobral15} using the data obtained 
with the Subaru Telescope by  \citet{matthee15}. 
Follow-up observations of CR7  using the F110W (YJ) and F160W (H) wide filters of the WFPC3 
instrument on board  the HST by \citet{sobral15}, revealed  
that CR7 is composed of three components, namely A, B and C. 
Most recently, \citet{sobral17} presented new Hubble/WFC3 grism observations
and a re-analysis of the VLT data of CR7, making it possible to obtain
line fluxes of the components A, B and C. In their analysis, \citet{sobral17}
  measured  the line \ion{N}{iv}]$\lambda$1483.4\AA, $\lambda$1486.6\AA\,
in clump A  (not considered in our analysis) and 
 \ion{N}{v}$\lambda$1240\AA\, in clump C. Although CR7 is the most studied among the objects in our sample, its nature is still uncertain (see Sect.~\ref{intro}).

\subsubsection{J233408}

\citet{shibuya18},  using the Subaru Hyper Suprime-Cam (HSC) survey
data,  measured the Ly$\alpha$ and \ion{C}{iv} fluxes and defined upper
limits for \ion{N}{v}, \ion{He}{ii} and \ion{O}{iii}] lines of this object.  These authors compared 
the \ion{He}{ii}/\ion{C}{iv} and \ion{O}{iii}]/\ion{C}{iv} ratios
 of this object  
 with those observed in a sample of  SFs and AGNs as well as with predictions of photoionization models 
 built by \citet{feltre16}. They  found that the constraints on these ratios are more compatible with
 SFs rather than AGNs.

\subsubsection{COSY}

\citet{laporte17} used  the XSHOOTER instrument on the  VLT
 and the MOSFIRE instrument on the Keck telescope 
 and obtained spectroscopic data of the
LAE COSY (see also \citealt{stark17}).  \citet{laporte17}   measured  the 
Ly$\alpha$, \ion{He}{ii} and one metal line, i.e. \ion{N}{v}$\lambda$1240\AA\
  (see Table~\ref{tab1}). From the comparisons of the ratios
of the observed emission-line fluxes \ion{C}{iii}]/\ion{He}{ii},
\ion{N}{v}/\ion{C}{iv},  \ion{C}{iii}]/\ion{He}{ii} and \ion{N}{v}/\ion{He}{ii}
with those predicted by photoionization
models built by \citet{nakajima17}, \citet{laporte17}
found that  COSY is a likely AGN host.

\subsubsection{A1703-zd6}

A1703-zd6 is a bright LAE, identified by \citet{bradley12} for which
first spectroscopic data was obtained by \citet{schenker12}.
Follow-up Keck/MOSFIRE observations  by \citet{stark15}
revealed the presence of UV metal emission-lines  in the spectrum of this object.  
To interpret the combined stellar and nebular emission of this object,
\citet{stark15}  found that  photoionization models can reproduce the observational data of A1703-zd6  if  a young stellar population together with  a gas component of metallicity  $(Z/Z_{\odot})\approx0.02$, or an AGN with $(Z/Z_{\odot})=0.001$ is assumed as the ionizing source.

 \begin{table*}
\caption{Emission-line fluxes for LAEs compiled from the literature. Flux of  \ion{O}{iii}]$\lambda$1666 
for A1703-zd6 was calculated assuming the  theoretical line flux ratio: \ion{O}{iii}]$\lambda$1660.81\AA/$\lambda$1666.15\AA=0.34.}
\vspace{0.3cm}
\label{tab1}
\begin{tabular}{@{}lccccccccc@{}}
\hline
\noalign{\smallskip}  
 Object                         &       redshift  &   Ly$\alpha$$\lambda$1216   &     \ion{N}{v}$\lambda$1239           &     \ion{C}{iv}$\lambda$1549            &            \ion{He}{ii}$\lambda$1640  & \ion{O}{iii}]$\lambda$1666   &       \ion{C}{iii}]$\lambda$1909          &  Flux ($\rm erg/s/ cm^{2}$)  &Ref.  \\
\noalign{\smallskip}
ID3                             &       6.110     &     $33.2\pm2.3$                   &         $ < 1.8$                                &        $14.0\pm3.8$                           &                        $<\:1.5$                &        $4.4\pm0.85$               &                  $<\: 3.6$                          &  $10^{-18}$                                 & 1  \\                   
                                   &                    &                                            &                                                      &                                                       &                                                    &                                            &                                                         &                                                    &     \\                          
CR7  C                       &       6.604     &     $2.6\pm1$                       &           $1.3\pm0.5$                         &           $<\:0.7$                                &                   $1.0\pm0.4$               &         $<\:0.8$                      &                   $<\:1.0$                           &    $10^{-17}$                                & 2   \\                          
                                   &                    &                                             &                                                    &                                                       &                                                     &                                           &                                                          &                                                     &      \\
J233408               &      5.707      &        $13.5\pm0.03$               &            $<\:0.67$                          &             $1.15\pm0.14$                     &              $<\:0.16$                       &        $<\:0.09$                      &                   ---                                   &     $10^{-17}$                                 & 3  \\ 
                                   &                     &                                             &                                                    &                                                       &                                                     &                                            &                                                       &                                                    &     \\
COSY                          &       7.149       &        $22.9\pm3.0$                 &             $2.58\pm0.44$                 &              $<2.70$                             &               $1.26\pm0.29$                &          ---                               &               $<\:1.75$                        &        $10^{-18}$                           & 4  \\
                                  &                       &                                             &                                                    &                                                       &                                                     &                                            &                                                       &                                                    &     \\
A1703-zd6                  &        7.045       &        $28.4\pm5.3$                 &            ---                                    &              $7.9\pm1.13$                     &               $<2.1$                            &           7.1                       &                   ---                                 &        $10^{-18}$                            & 5 \\         
\noalign{\smallskip}
\hline
\end{tabular}
\begin{minipage}[c]{2\columnwidth}
References---  Data taken from   (1) \cite{mainali17}, (2) \citet{sobral17},      
(3) \citet{shibuya18},  (4) \citet{laporte17}, \\ and 
(5) \citet{stark15}.
\end{minipage}
\end{table*}


\subsection{Control Sample}
\label{control}

 We compiled   observational UV narrow emission line intensities  ($\rm FWHM \: \la \: 1000$ km/s)
of a sample of   SFs and AGNs from the literature.  Their emission-line intensity ratios were compared with those of the selected LAEs.

Concerning the SFs, we considered {13 objects} at redshift $z\: < 3.4$. This sample consists of 
8  objects (\ion{H}{ii} regions and  starburst galaxies) compiled by \citet{perez17} {whose data were}
obtained   by \citet{villar04}, \citet{erb10}, 
\citet{berg16},  \citet{debarros16},  \citet{steidel16}, and \citet{vanzella16}. {For the remaining five star forming galaxies, the data of one (object M0451) was obtained from \citet{stark14}
and four (objects SB2, SB82, SB111, SB182) from \citet{senchyna17}.}
Regarding  the AGN sample, we considered the observational data of 80 objects 
compiled by \citet{dors14}, being 11 Seyfert 2 nuclei ($z \: < \:0.04$), 
10 type 2 Quasars ($1.5 \: < \: z \: < \: 3.7$) and 59 high-$z$ radio galaxies 
($1.2 \: < \: z \: < \: 3.8$). The reader is referred to \citet{dors14} for references pertaining to the original AGN sample.

\section{Photoionization models}
\label{mod}
 
We used the {\sc Cloudy} code version 17.00 \citep{ferland13} to build a grid of photoionization models 
in order to compare the predicted UV emission-line intensities   with those measured for the
LAE spectra (see Sect.~\ref{dlae}).

The models were built assuming a simple approach, i.e. a static  spherical  geometry with a central ionizing source, and constant electron density along the radius with different values of metallicity  and number of ionizing photons
 emitted per second by the ionizing source.  
The gas ionization was assumed to be due to  photoionization from radiation emitted by a unique source, and we did not consider gas shock
ionization and heating. This is a good assumption for LAEs because the majority of these objects exhibit
narrow emission lines, hinting that gas shock waves are absent, or  have only a small influence
on the ionization/heating of the gas.  We assumed three distinct types for the ionizing sources: (i) a power law to represent Active  Galactic Nuclei, (ii) 
Direct Collapse Black Hole and (iii)  Population II stars.  
In what follows, the main parameters of the models are described.

\subsection{Active Galactic Nuclei}
\label{sagn}
To simulate the observational properties of AGNs, we considered the following values for
the metallicity (Z) in relation to the solar one ($Z_{\odot}$): 
 $(Z/Z_{\odot})$=
0.01, 0.5, 1.0 and 2.0.
These values cover the range of metallicities derived for AGNs  located  in a wide range of redshift  ($0.0 \: < \: z \: < \: 4.0$)
as derived, for example, by \citet{dors14, dors15, dors17a}, \citet{castro17}, and \citet{groves06}, who compared
 photoionization model results  with observational narrow  UV and optical emission line intensities 
of AGNs (see also \citealt{revalski18, thomas18}).

The abundance of the heavy elements was scaled with the oxygen abundance, with the exception of the nitrogen abundance,
which was derived from the relation obtained by \citet{dors17a}:
\begin{equation}
\rm \log(N/H)=1.05 (\pm0.09) \times [\log(O/H)] -0.35 (\pm 0.33).
\end{equation}
{The authors derived this relation  using the {\sc Cloudy}  code to build detailed
photoionization models  to reproduce observational optical narrow emission line
intensities of 44 Sy2 AGNs at $z\: < \:0.1$ compiled from the literature.}

\begin{figure}
\centering
\includegraphics[angle=-90,width=1\columnwidth]{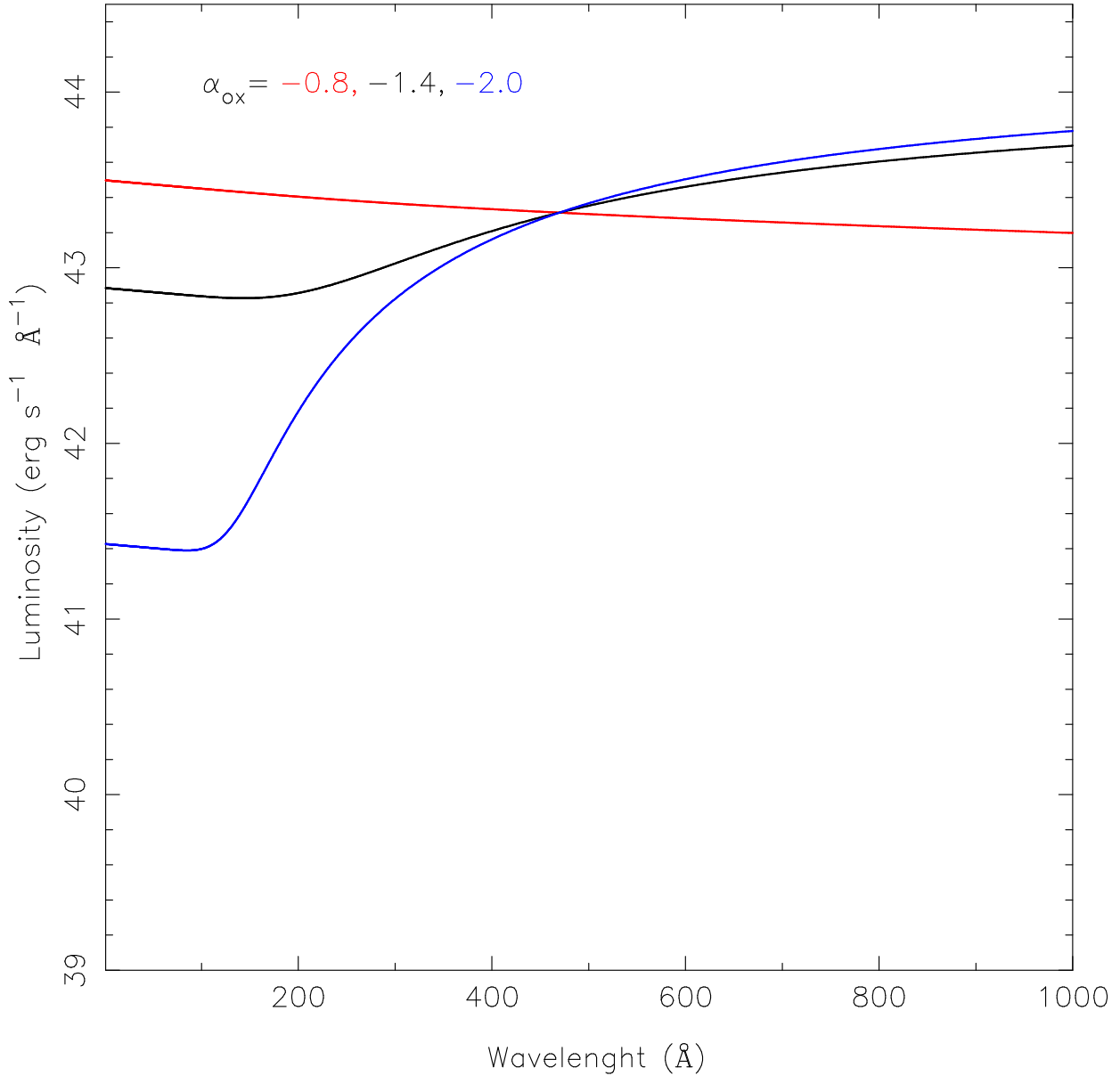}
\caption{Rest-frame SED (logarithm of the luminosity vs.\ wavelength) in the UV spectral region 
($\lambda \le \rm \: 1000 \: $\AA) used in  the AGNs photoionization models considering three 
$\alpha_{ox}$ values, represented with distinct colours,  as indicated. The logarithm of the number of ionizing photons
considered is $\log Q(\rm H)=54$.}
\label{f1}
\end{figure}

For the models, we assumed an electron density value of $N_{\rm e}=500 \: \rm cm^{-3}$, which is  
 a representative value for the gas density of  Narrow Line Regions  as found by  \citet{dors14}. 
The SED  was considered to be composed of two continuum components,
where one represents the Big Blue Bump peaking at $\rm 1 \: Ryd$, and the other is characterized by a power law with 
 different  values of the spectral index $\alpha_{ox}$  describing the continuum between 2 keV and 2500\AA  \, \citep{zamorati81}.
In the models, we consider the $\alpha_{ox}$ values $-$0.8, $-1.4$ and $-2.0$, in order to cover 
the entire range of observed values derived by \citet{miller11}. In Figure~\ref{f1} we show the SEDs 
assuming these $\alpha_{ox}$ values, with logarithm of the number of ionizing photons emitted per 
second by the ionizing source $\log Q(\rm H)=54$ and for
$\lambda \: \leq \:1000 $ \AA.

The distance from the ionizing source to the innermost region of the gas ($R_{\rm int}$) was assumed to be 3 pc,
which is similar to the radius derived in observational studies for narrow line regions (e.g. \citealt{jaffe04, lopez16}).
The range of values for the number of ionizing photons  was considered to be
$50 \: \leq \: \log Q(\rm H) \: \leq 55$ with  a step of 1 dex.  In total, 90 photoionization models were built.

 \subsection{Direct Collapse Black Hole}
 \label{sdcbh}
 
 For these models,  the same AGN metallicity,  $R_{\rm int}$, $N_{\rm e}$
 and $Q$(H) values were assumed. The SEDs  were considered to be 
 multi-temperature black body spectra as defined by  \citet{shakura73}, which simulate the
  radiation being emitted from a black hole accreting gas. We consider four distinct parameters for this system: 
  black holes with masses of $10^{6} \: M_{\odot}$ and $10^{7} \: M_{\odot}$ accreting at $50\%$ and $100\%$ of the Eddington rate.
 These are the same SEDs considered in the photoionization models built  by  \citet{agarwal17}. 
 
\begin{figure}
\centering
\includegraphics[angle=-90,width=1\columnwidth]{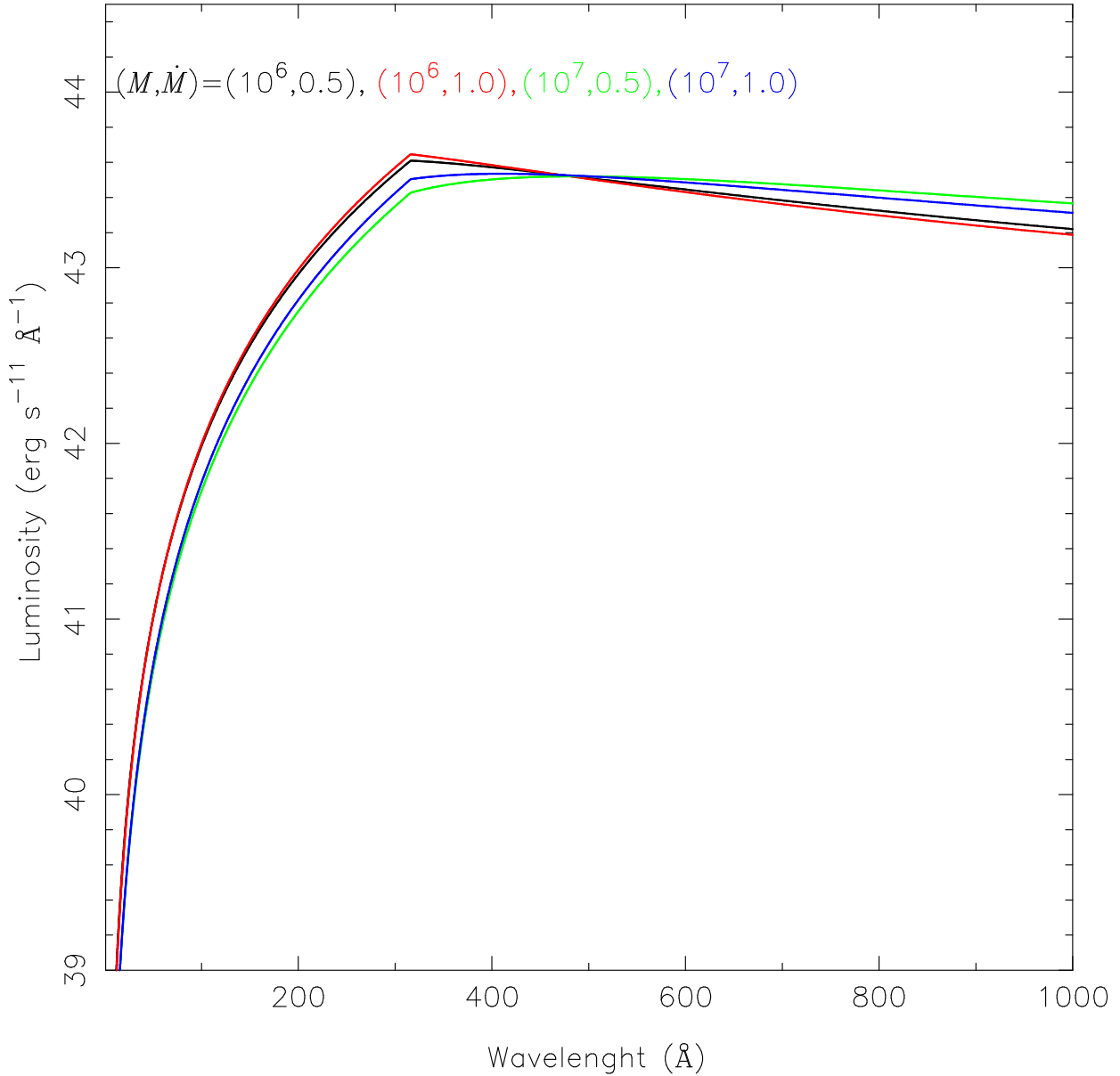}
\caption{Similar to Fig.~\ref{f1} but for the SEDs assumed in the DCBH models.  Curves  represent
  multi-temperature black body, whose
  radiation arises from a black hole with different masses and accreting gas at different rates, as indicated.
 The mass $M$ is in unit of solar masses and $\dot{M}$ is in units of Eddington rate.}
\label{f2}
\end{figure}

The theory predicts that DCBHs form in an environment that is nearly metal-free  in order to prevent fragmentation
into stars \citep{bromm03, omukai08, latif16}. {However, the host haloes of these DCBHs quickly merge with a neighbouring galaxy and, therefore, the enriched 
gas in our models is a consequence of the later evolution of the DCBH, as proposed by \citet{agarwal13}}. 
In Fig.~\ref{f2} the SEDs assumed for the
 DCBH models are presented. 
We can see that SED models with the same BH mass are very similar, implying that the mass 
accretion rate has a small influence on the SED.
 The model considering   $10^{6} \: M_{\odot}$ is somewhat harder than the one for
 $10^{7} \: M_{\odot}$. A total of 95 models were built.

 \subsection{Population II stars}
 \label{spopii}
 
 In the models for Population II (PopII) stars, the gas metallicity  values were considered to be $(Z/Z_{\odot})$=0.03, 0.2, 0.5, and 1.0, 
 about the same  range of values found in \ion{H}{ii} regions located in  the disks of nearby spiral galaxies
 (e.g.\ \citealt{kennicutt03, Diaz+07, dors17b}) and in star-forming galaxies (e.g. \citealt{kewley08,Hagele+06,hagele08,Hagele+11,Hagele+12}).
 The  electron density value  was considered to be $N_{\rm e}=200 \: \rm cm^{-3}$   in the models, 
 which  is in the range of those derived from the  [\ion{S}{ii}]$\lambda$6716/$\lambda$6731
  emission lines ratio for extragalactic \ion{H}{ii} regions \citep{copetti00, krabbe14, sanders16}.

The abundances of  heavy elements were linearly scaled with the oxygen abundance, with  exception of N
and C. For nitrogen, we used the following  relation between this element and oxygen derived by \citet{vilas93}:
\begin{equation}
\label{eq1a}
\rm 
\log{N/H}=\log(O/H)+\log(0.034+120\: \times \:O/H).
\end{equation}
{Regarding carbon,  there are hints suggesting that galaxies located at high redshift ($z=2-4$) have an enhanced C/O abundance ratio
in comparison to the objects at low redshift \citep{nakajima17}.  Since the C/O-O/H abundance relation 
assumed in the models affects the predicted strengths of \ion{C}{iii}] and \ion{C}{iv} lines \citep{nakajima17},
we consider two relations in our grid of models.}  Firstly, a grid of photoionization models was built considering $\rm \log(C/O)$ equal to the solar ratio
$\rm \log(C/O)_{\odot}=-0.30$ obtained by \citet{allende-prieto01, allende-prieto02}. Alternatively, another grid of models
was built  assuming  the following derived by \citet{dopita06}:
 \begin{equation}
 \label{eq2}
\rm  C/H=6.0 \: \times 10^{-5} \times (Z/Z_{\odot}) + 2.0 \: \times 10^{-4} \times (Z/Z_{\odot})^{2}.
 \end{equation}

\begin{figure}
\centering
\includegraphics[angle=-90,width=1\columnwidth]{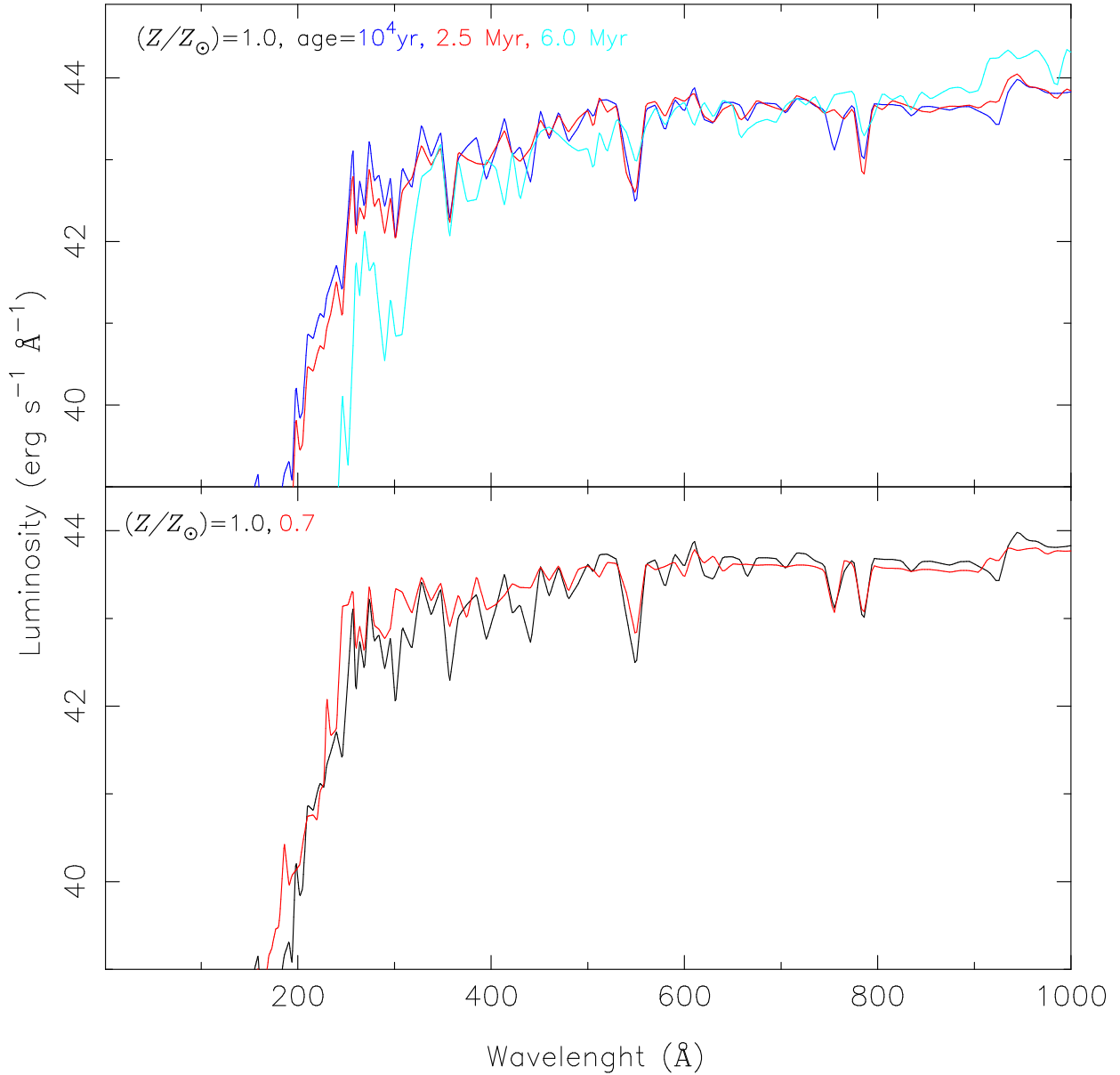}
\caption{Rest-frame SEDs of PopII stellar clusters built  using the {\sc  STARBURST99} code \citep{leitherer99}
and the GENEVA stellar tracks with rotation \citep{levesque12}. In the upper panel we compared SEDs with Z/Z$_\odot$=1.0 and different ages as labelled. In
  the lower panel we show a comparison between SEDs built assuming an age of 0.01\,Myr and two different metallicities. 
  In all cases the WM-basic stellar atmospheres \citep{paudrach01} were assumed.
}
\label{f3}
\end{figure}

As a first step, the ionizing source was considered to be a stellar cluster that formed instantaneously (i.e burst) with a mass of  $10^{6} \: M_{\odot}$, following an Initial Mass Function (IMF) with exponents of 1.3 for 
 stellar masses from 0.1 to 0.5 $M_{\odot}$ and 2.3 for stellar masses between 0.5 
 and 100 $M_{\odot}$.
These parameters were assumed as our input for the {\sc  STARBURST99} code \citep{leitherer99} in order to 
generate the SEDs. {Two stellar tracks were considered in the  {\sc  STARBURST99} models:
(i) the GENEVA  tracks with stellar rotation \citep{levesque12} with metallicities
of 0.7 and 1.0 $Z/Z_{\odot}$ (the only available metallicities) and (ii) 
the PADOVA tracks with AGB stars \citep{bertelli94} with metallicities
of 0.02, 0.2, 0.4, and 1.0 $Z/Z_{\odot}$.  }
Three ages were considered for the stellar cluster: $10^{4}$\,yr, 2.5\,Myr, and 6.0\,Myr. 
Theoretical studies based on comparison 
between emission-line intensities of \ion{H}{ii} regions predicted 
by photoionization models  and  observational data  have found  
ionizing stellar cluster   with ages
in this range  \citep{copetti85, bresolin99, stasinska03, dors06}.
We consider the  WM-basic stellar atmosphere models of  
 \citet{paudrach01}, which  seem to reproduce the best agreement between predicted and observed optical
emission-line ratios of \ion{H}{ii} regions \citep{zastrow13}. {In
  the upper panel of Fig.~\ref{f3}, the GENEVA stellar cluster SEDs
  with different ages and the same metallicity (Z/Z$_\odot$=1.0) are
  shown. In the bottom panel of the same figure, a comparison between the 
  stellar clusters SEDs from GENEVA assuming an age of  0.01\,Myr and two different metallicities (Z/Z$_\odot$=1.0 and  Z/Z$_\odot$=0.7) are shown. The dependence of the SED on both the age and metallicity can be seen in the figure (see also  \citealt{senchyna17, leitherer99}).}

Thus, we obtained {18 different SED models with $\log Q(\rm H)$ varying between $\sim$\,51.1 and $\sim$\,52.8.} 
 In the second step we built the complete PopII photoionization models by varying the logarithm  of 
the number of ionizing photons  in the 
range $\rm 49 \: \leq \: \log Q(H) \: \leq 56$, with a step of 1 dex.
Since the number of ionizing photons is directly proportional to the stellar cluster mass,
this  $Q(\rm H)$  range, considering the {\sc  STARBURST99} \citep{leitherer99} predictions,
can be obtained by varying the stellar cluster mass in the range $\sim 10^{4}$ to $\sim 10^{11}$ $M_{\odot}$. {In order to build realistic models, the metallicity of the gas phase was matched with the closest available metallicity of
the stellar atmospheres (see \citealt{dors11} for a discussion about this methodology). 
In total, 272 models were built.}

\medskip

\begin{figure}
\centering
\includegraphics[angle=-90,width=1\columnwidth]{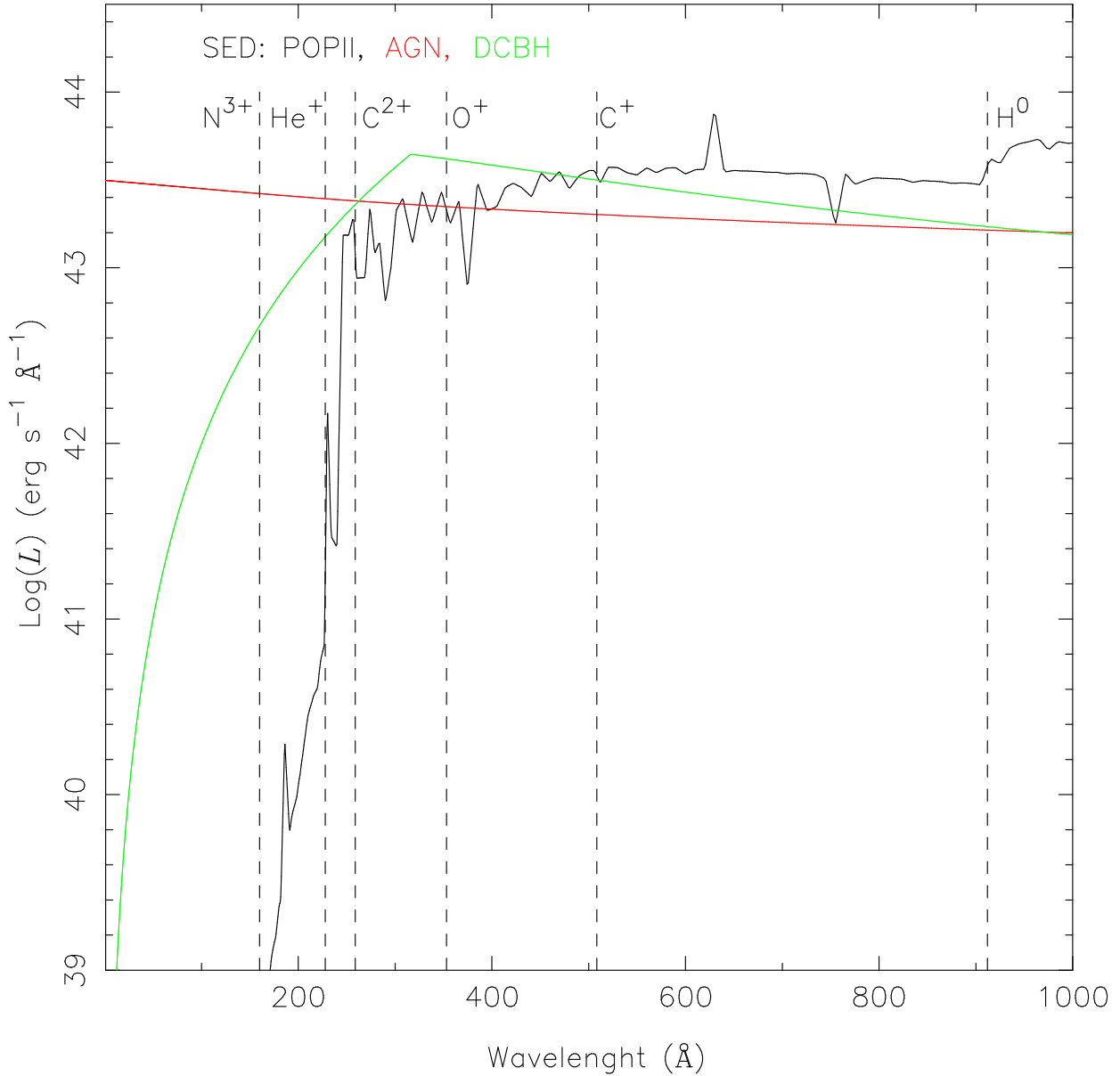}
\caption{Comparison among the hardest SED of each type considered in the models (as indicated) and described 
in Sect.~\ref{mod}. Dashed lines mark the wavelengths corresponding to the ionization potential 
of {H$^0$, C$^+$, O$^+$, C$^{2+}$, He$^{+}$ and N$^{3+}$. The hardest SED for PopII stars corresponds to the one with the lowest metallicity
($Z/Z_{\odot}=0.02$) and 0.01 Myr old,  for AGNs it corresponds to the one with $\alpha_{ox}=-0.8$ and 
for DCBH  it corresponds to the one with a mass of $10^{6} \: M_{\odot}$ accreting 100\% at Eddington rate.}}
\label{f5}
\end{figure}

In Fig.~\ref{f5} we present a comparison between the hardest SED of each type, as described  previously. 
 It shows that our AGN model exhibits the hardest SED, followed by the DCBH and PopII model.

\begin{table*}
\caption{Indication of which type of model grid (see Section~\ref{mod})
reproduces the observed emission line ratios of the sample of objects in the 
diagrams presented in Fig.~\ref{fr5}.
Between parentheses, the alternative classification that could be derived if we take into account the arrows (added when only upper or lower limits of 
the line ratios are available) and the intrinsic errors in the line-curves determinations are indicated.} 
\vspace{0.3cm}
\label{tab2}
\begin{tabular}{@{}lccccc@{}}
\hline
\noalign{\smallskip}  
                                                                &       \multicolumn{5}{c}{Classification}                                                                                                             \\
\noalign{\smallskip}						
                                                               &                     ID3                   &      CR7(C)               &   J233408   & COSY  & A1703-zd6     \\
  							       
\noalign{\smallskip}  							       
							       
	Diagram						  &                                                                       &                                   &                  &              &                      \\
\cline{1-6}
\noalign{\vspace{2pt}}	
 	                                                               &                   \multicolumn{5}{c}{Empirical}                                                                                                      \\ 
 \noalign{\smallskip}
\ion{C}{iii}]/\ion{He}{ii} vs. \ion{N}{v}/\ion{He}{ii}    &                           ---                                          &         AGN                 &      ---        &   AGN     &      ---            \\
\ion{C}{iii}]/\ion{He}{ii} vs. \ion{C}{iv}/ \ion{He}{ii}   &                           ---                                          &         AGN                 &      ---        &   AGN     &      ---          \\
C43  vs.  \ion{C}{iii}]/\ion{C}{iv}                             &                          SF                                         &         ---                      &    ---         &   AGN     &     ---               \\
\ion{C}{iii}]/\ion{He}{ii}  vs. \ion{O}{iii}]/\ion{He}{ii}  &                         ---                                           &         SF (AGN)           &     ---        &    ---        &     ---               \\
\ion{C}{iv}/Ly$\alpha$ vs.\ion{O}{iii}]/Ly$\alpha$     &                        AGN (SF)                      &         SF (AGN)            &     AGN     &   ---         &     SF                    \\
\cline{1-6}
\noalign{\vspace{2pt}}

 	                                                                &                   \multicolumn{5}{c}{Theoretical}                                                                                                    \\ 
 \noalign{\smallskip}
\ion{C}{iii}]/\ion{He}{ii} vs. \ion{N}{v}/\ion{He}{ii}  &                         ---                                          &         AGN                  &       ---        &    AGN       &   ---           \\
\ion{C}{iii}]/\ion{He}{ii} vs. \ion{C}{iv}/ \ion{He}{ii} &                         ---                                         &          AGN                  &       ---         &   AGN        &  ---            \\
C43  vs.  \ion{C}{iii}]/\ion{C}{iv}                           &                        AGN (SF)                                &         ---                       &      ---         &    AGN        &   ---           \\
\ion{C}{iii}]/\ion{He}{ii}  vs. \ion{O}{iii}]/\ion{He}{ii} &                         ---                                        &          AGN                    &    ---           &    ---         &    ---          \\
\ion{C}{iv}/Ly$\alpha$ vs.\ion{O}{iii}]/Ly$\alpha$    &                        SF (AGN)                               &          SF (AGN)             &     AGN        &   ---          &     SF      \\
\hline
Adopted classification                                          &                          SF                                       &          AGN                   &     AGN       &   AGN      &   SF           \\             
\noalign{\smallskip}
\hline 
\end{tabular}
\end{table*}

\section{Results}
\label{res}

\subsection{Spectral Classification}

 \begin{figure}
\centering
\includegraphics[angle=-90,width=1\columnwidth]{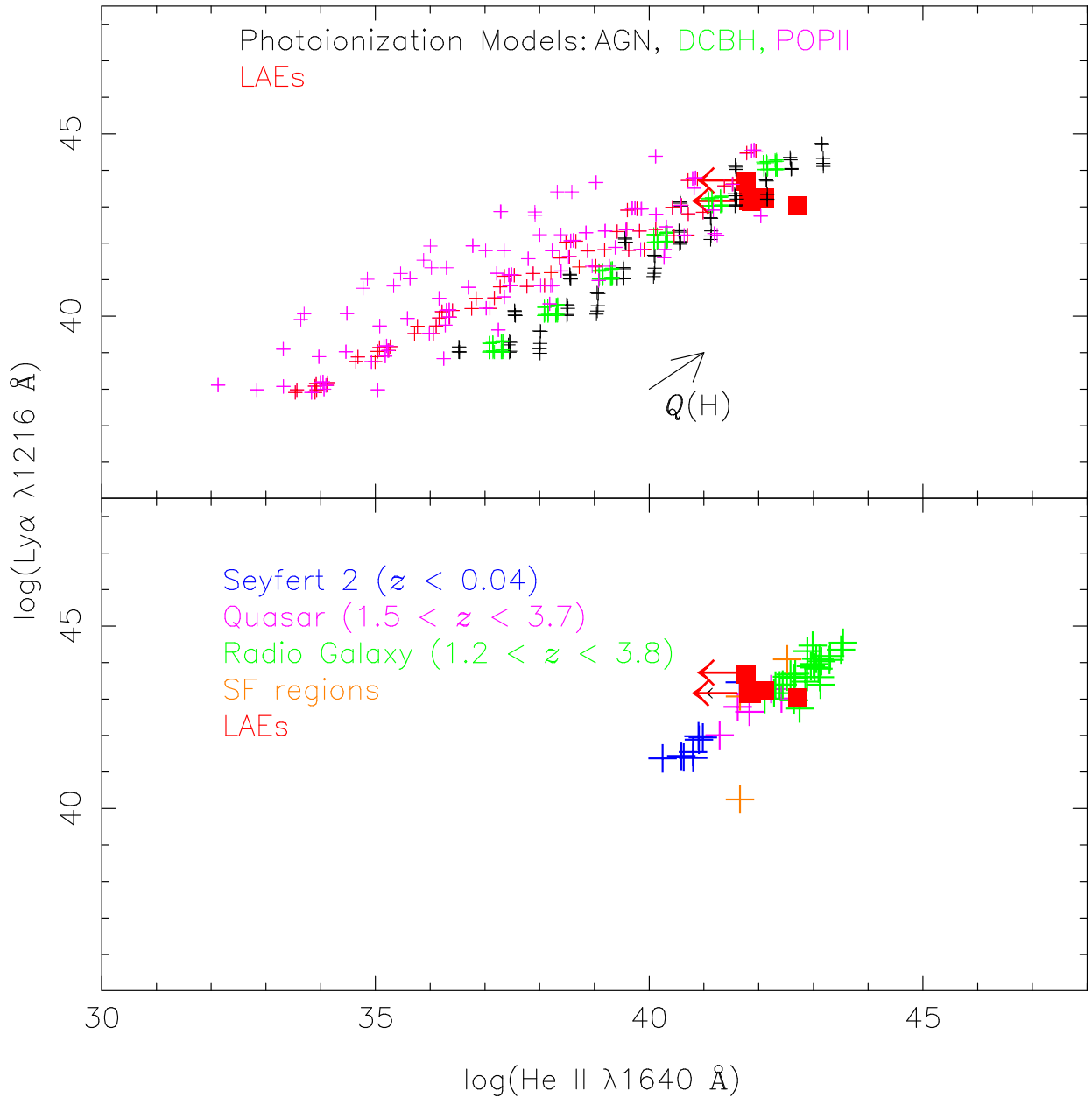}
\caption{Logarithm of the $\rm Ly\alpha$ luminosity versus 
the logarithm of $\rm \ion{He}{ii}\: \lambda1640 \:\AA$ luminosity. Bottom panel: Comparison
between the luminosities of the objects of the control sample (see Section~\ref{control}) and  those of the LAEs  listed in Table~\ref{tab1}. Upper panel:  Comparison 
between the luminosities predicted by the models (see Sect.~\ref{mod}) and of the LAEs.
The black arrow indicates the direction in which the number of ionizing photons increases in the models.
The meaning of each symbol is indicated in the plots. The arrows linked to the object symbols indicate
that only the upper limit of the \ion{He}{ii} luminosity values  were quoted. {Typical errors in the 
luminosity values are about 0.1 dex \citep[see e.g.][]{matthee17, shibuya18}. }}
\label{fr1}
\end{figure} 

In order to spectroscopically classify the five selected LAEs, we took into account
 diagnostic diagrams containing observational line ratios of the control sample and
the ones predicted by the photoionization models, and created
demarcations between the regions occupied by SFs and AGNs.

In Fig.~\ref{fr1}, bottom panel, we compared the logarithm of the luminosity of
Ly$\alpha$ and \ion{He}{ii} calculated for the LAEs with the ones of the control sample.  
The luminosity of each object was calculated using the fluxes listed in Table~\ref{tab1} and
 the fluxes obtained from the papers from which the control sample data were compiled.
The distance to each object was calculated using  its redshift and  assuming a spatially flat cosmology with
 $H_{0}$\,=\,71 $ \rm km\:s^{-1} Mpc^{-1}$, $\Omega_{m}=0.270$, and $\Omega_{\rm vac}=0.730$  \citep{wright06}. 
Unfortunately, in Fig.~\ref{fr1}, it was possible to calculate
the Ly$\alpha$  and \ion{He}{ii} luminosities only for three SFs (the objects observed by \citealt{debarros16, steidel16, erb10}). From this plot, one can conclude the following:
\begin{enumerate}
\item  A trend is apparent. 
\item LAEs show luminosities similar to those seen for Quasars and less bright \subitem Radio Galaxies.
\item  The Ly$\alpha$ luminosities of the LAEs are spread in a range of about a factor of 5, \subitem while the \ion{He}{ii} ones present variations of about an 
order of magnitude, being relatively \subitem concentrated if we compared their luminosities with those of the objects in the\subitem control sample.
\item  LAE luminosities are higher than those of Seyfert 2 nuclei (except for the most \subitem luminous Seyfert 2 object).
\item More  data for SF is necessary to improve the comparison between\subitem these objects and LAEs.
\end{enumerate}

In the upper panel of Fig.~\ref{fr1}  we compare the LAE  luminosities
with those predicted by  our grid of photoionization models (see Sect.~\ref{mod}).
We found that model results describe  the LAE data well only when
values of $Q(\rm H)$ higher than 53 dex (value not indicated in the plot) are considered. 
Furthermore, no clear distinction among models built using different kind of ionizing sources can be made.

\begin{figure*} 
\centering
\includegraphics[angle=-90,width=0.42\columnwidth]{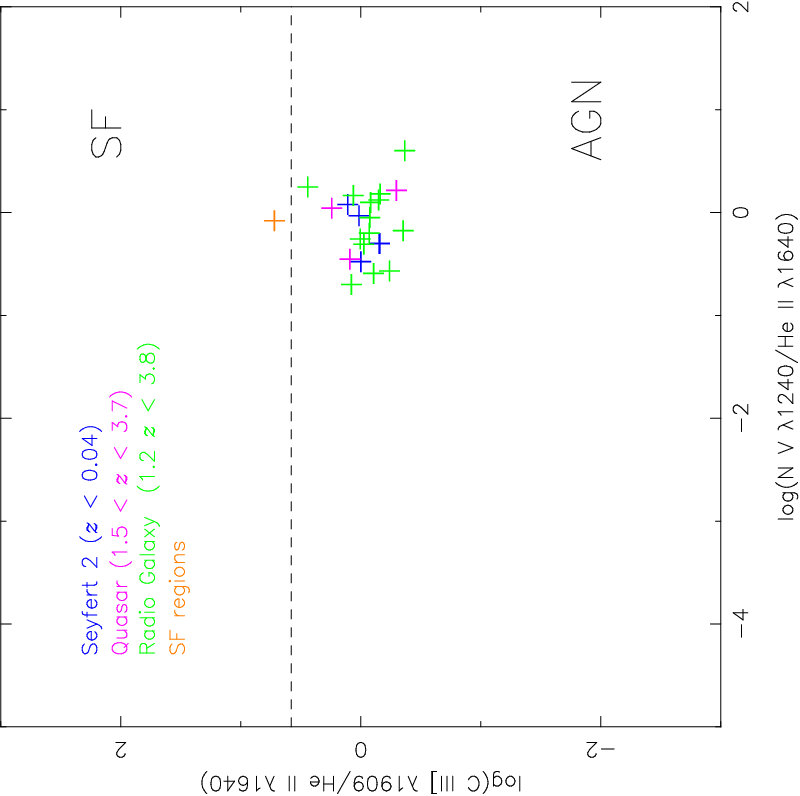}\hspace{0.05\columnwidth}
\includegraphics[angle=-90,width=0.42\columnwidth]{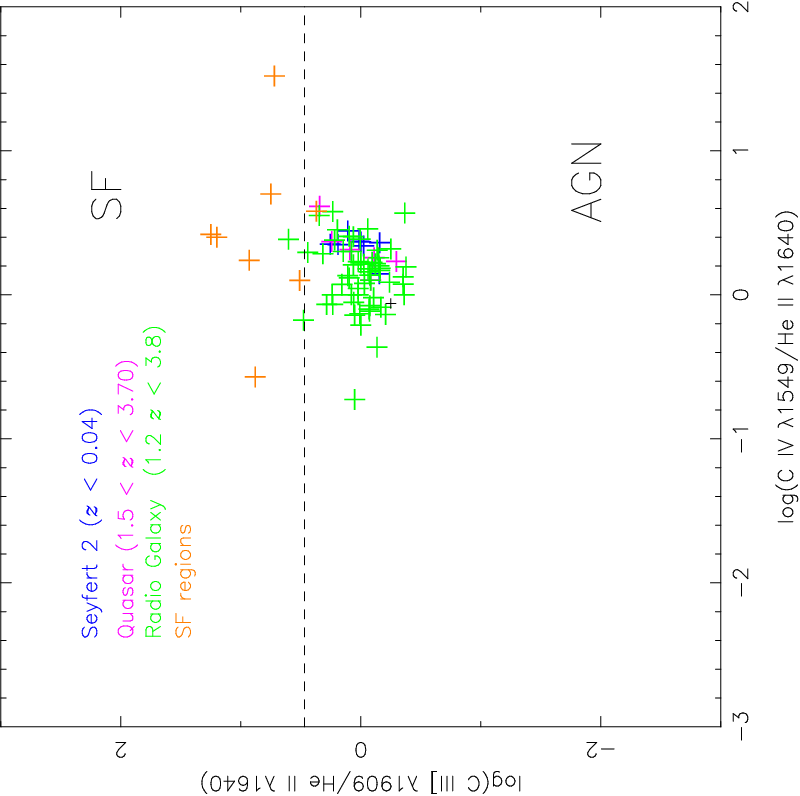}\vspace{0.05\columnwidth}
\includegraphics[angle=-90,width=0.42\columnwidth]{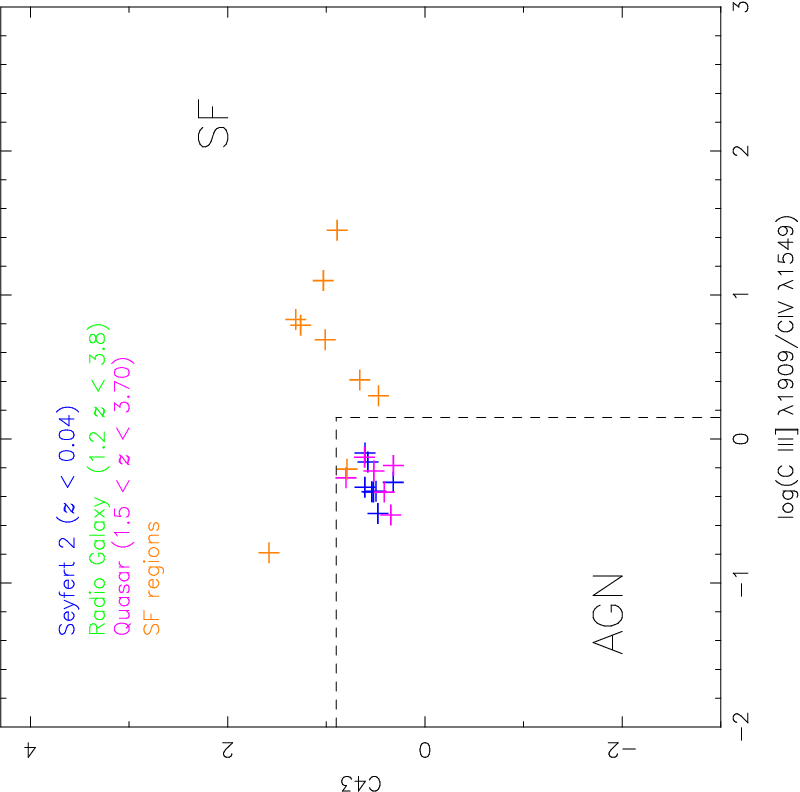}\hspace{0.05\columnwidth}
\includegraphics[angle=-90,width=0.42\columnwidth]{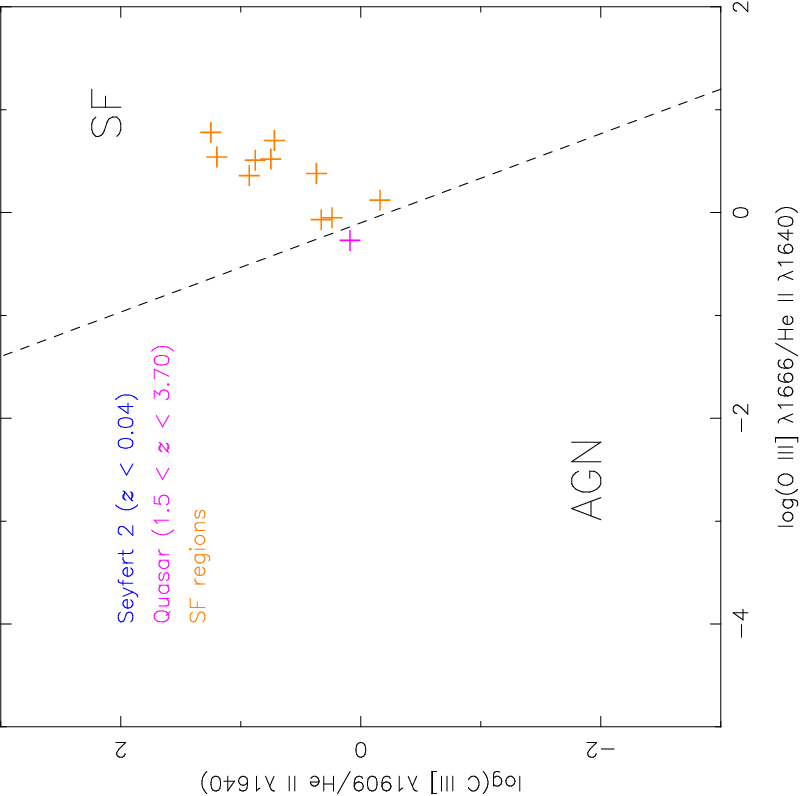}\vspace{0.05\columnwidth}
\includegraphics[angle=-90,width=0.42\columnwidth]{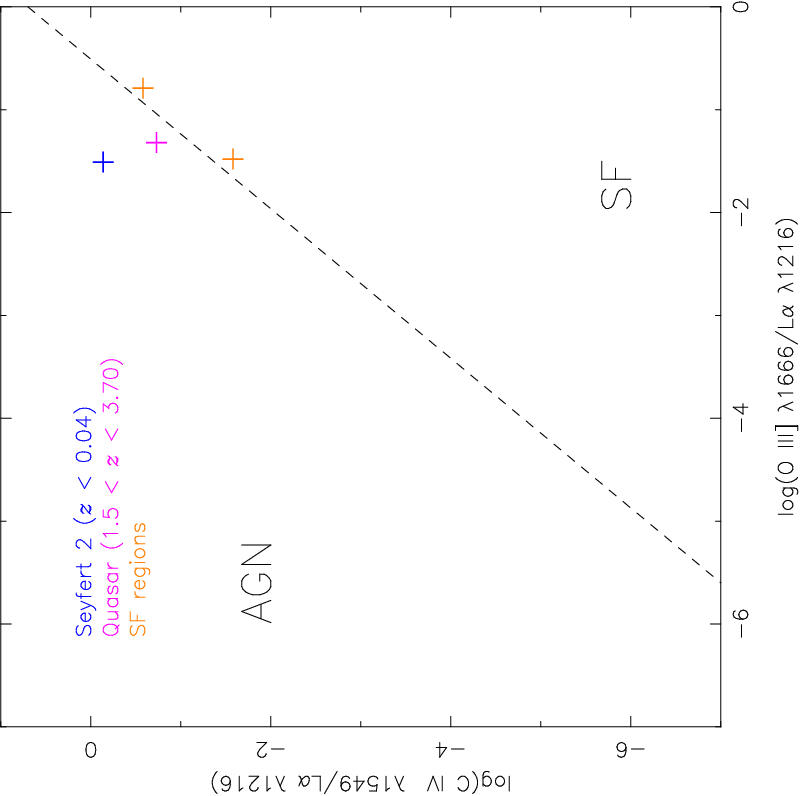}\hspace{0.05\columnwidth}
\includegraphics[angle=-90,width=0.42\columnwidth]{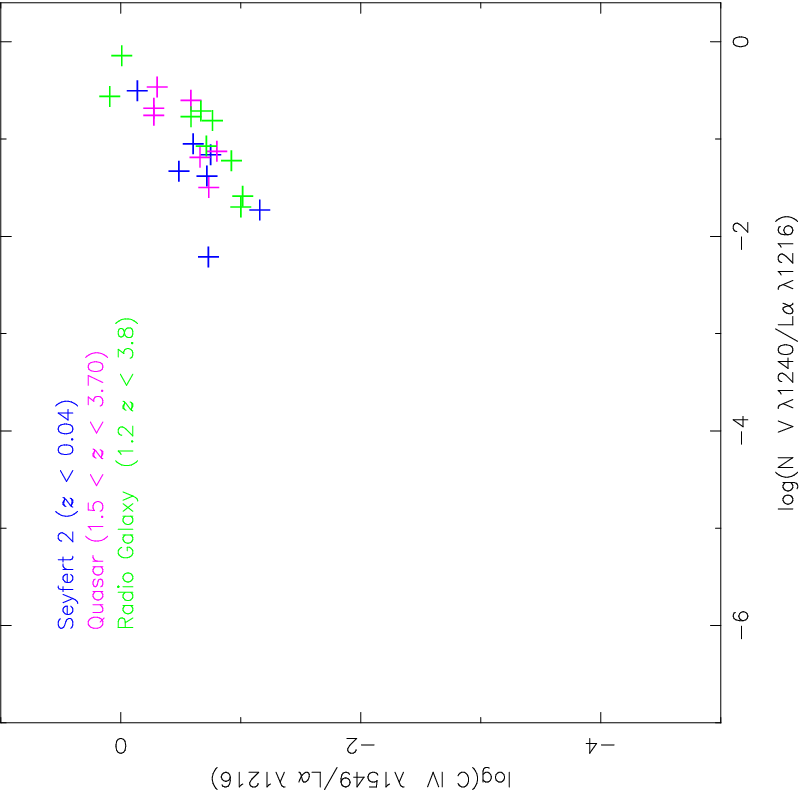}\vspace{0.05\columnwidth}
\caption{Diagnostic diagrams considering different emission-line ratios for the control sample (see Section~\ref{control}).
 The C43 refers to  log[(\ion{C}{iv}$\lambda$1549+\ion{C}{iii}]$\lambda$1909)/\ion{He}{ii}$\lambda$1640] \citep{dors14}.  
The line curves separate SF-like objects zone from that of AGN/DCBH-like objects. Symbols with different colours 
 represent   distinct object types as indicated.} 
\label{fr2}
\end{figure*}

In Fig.~\ref{fr2}, six diagnostic  diagrams  considering different emission-line ratios for the control sample are shown. 
In the \ion{C}{iv}/Ly$\alpha$ versus \ion{N}{v}/Ly$\alpha$ diagram (bottom-right panel) it was not possible to 
include SF due to lack of the 
involved emission-lines. Objects with only upper limits of the emission-line ratio are not included in the diagrams.
Regardless of the limited SF data, we  define demarcations between  the zones occupied by SFs and AGNs 
in all diagrams, with the exception of the \ion{C}{iv}/Ly$\alpha$ vs.\ \ion{N}{v}/Ly$\alpha$ diagram. {It is worth mentioning that due to  the fact the Ly$\alpha$ is a resonant
transition and is highly scattered by the neutral hydrogen in the Intergalactic Medium at high redshifts ($z \: > \: 5$),  
the Ly$\alpha$ flux is usually observed at about 5-10\% of its intrinsic output.  
The effect of a Ly$\alpha$ attenuation in the \ion{C}{iv}/Ly$\alpha$ vs.\ \ion{O}{iii}]/Ly$\alpha$ diagram 
would result in these objects moving almost parallel to the curves that separate SF from AGNs, since Ly$\alpha$ is in the denominator of both the axes.}

\begin{figure*} 
\centering
\includegraphics[angle=-90,width=0.42\columnwidth]{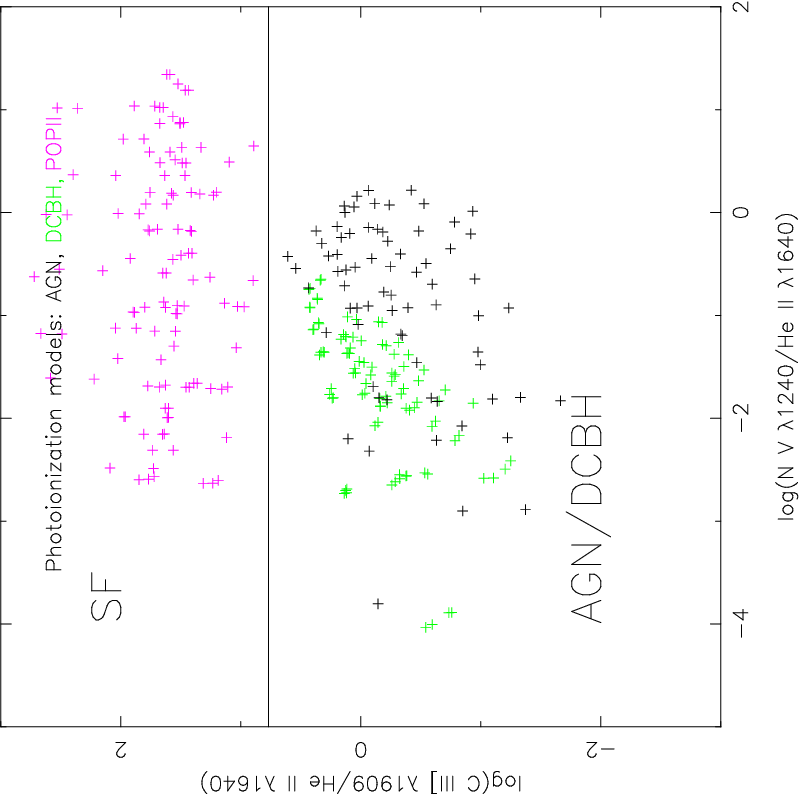}\hspace{0.05\columnwidth}
\includegraphics[angle=-90,width=0.42\columnwidth]{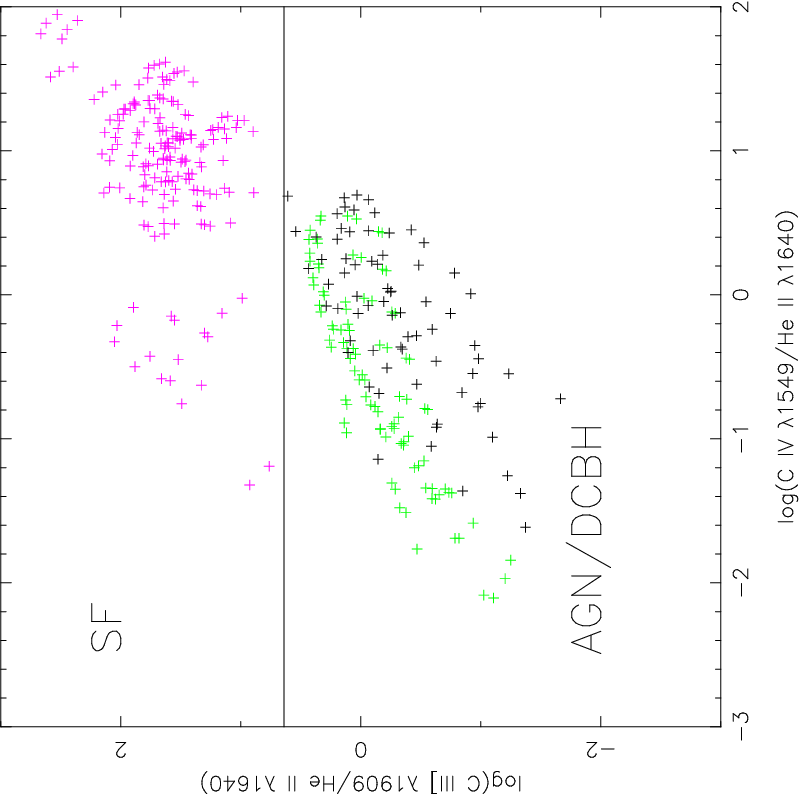}\vspace{0.05\columnwidth}
\includegraphics[angle=-90,width=0.42\columnwidth]{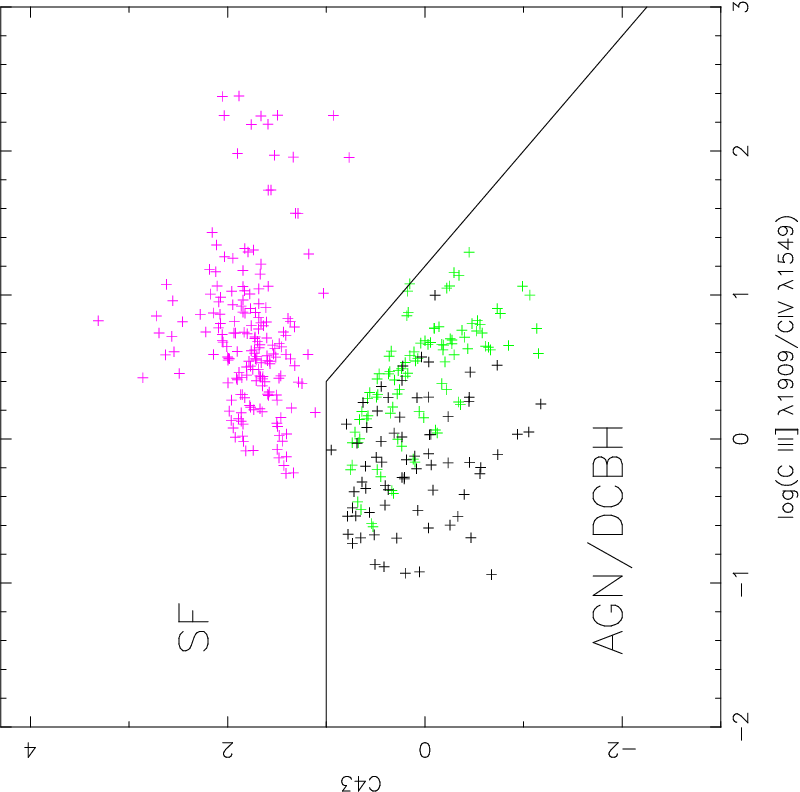}\hspace{0.05\columnwidth}
\includegraphics[angle=-90,width=0.42\columnwidth]{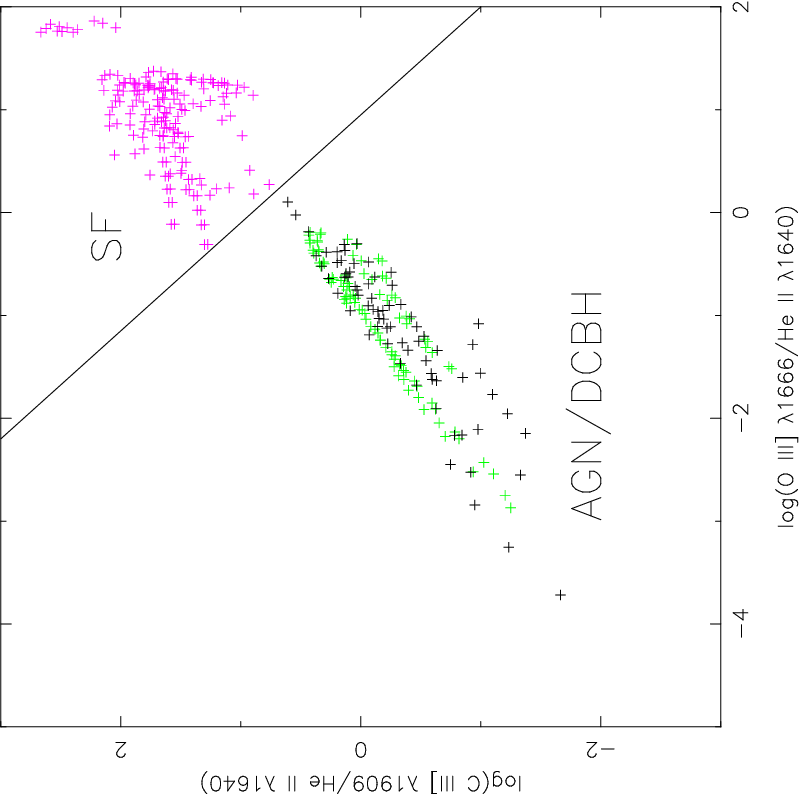}\vspace{0.05\columnwidth}
\includegraphics[angle=-90,width=0.42\columnwidth]{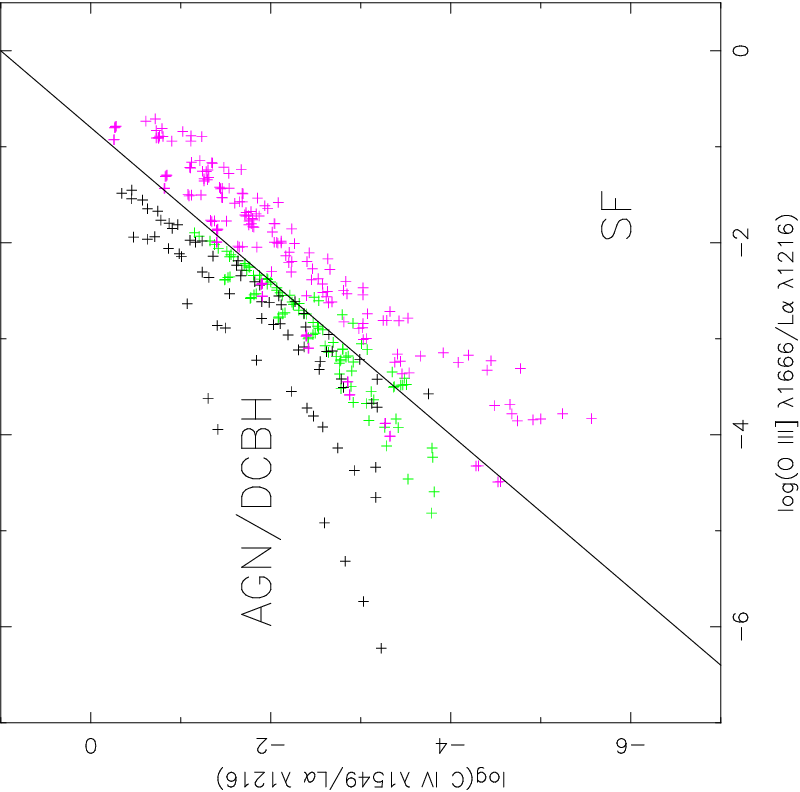}\hspace{0.05\columnwidth}
\includegraphics[angle=-90,width=0.42\columnwidth]{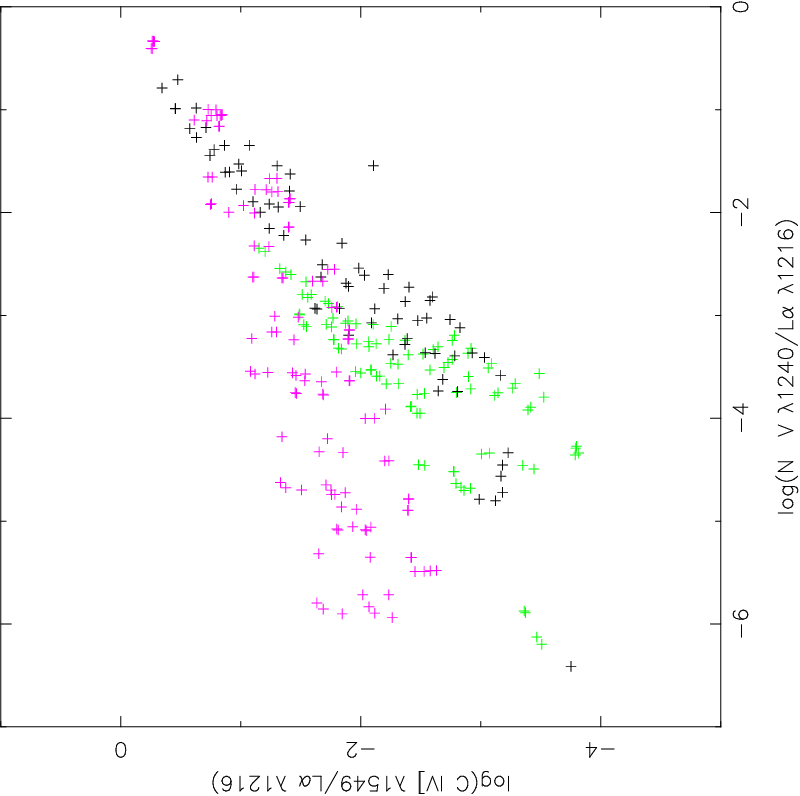}\vspace{0.05\columnwidth}
 \caption{Same as Fig.~\ref{fr2} but considering line ratio intensities predicted
 by the photoionization models (see Sect.~\ref{mod}). Symbols with different colours 
 represent models with distinct ionizing source as indicated.} 
\label{fr4}
\end{figure*}

The same set of diagnostic diagrams are also built considering our photoionization model results, as shown in Fig.~\ref{fr4} (see also \citealt{feltre16, nakajima17, nakajima18, sobral18}). 
In these diagrams, we can see that AGN and DCBH models occupy the same region, 
that is clearly different from the region occupied by the SF models, with exception of the 
\ion{C}{iv}/Ly$\alpha$ vs. \ion{N}{v}/Ly$\alpha$ diagnostic diagram where the different models overlap.
Demarcations between the zones occupied by SFs and AGNs were also defined. These zones are different from the ones empirically found using the control sample.
 
\begin{figure*} 
\centering
\includegraphics[angle=-90,width=0.42\columnwidth]{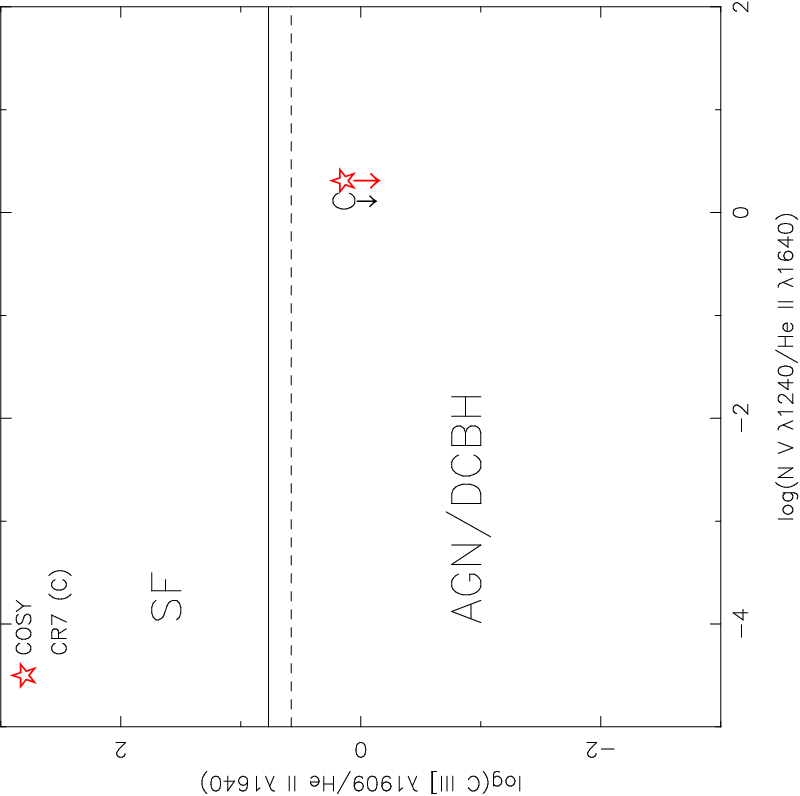}\hspace{0.05\columnwidth}
\includegraphics[angle=-90,width=0.42\columnwidth]{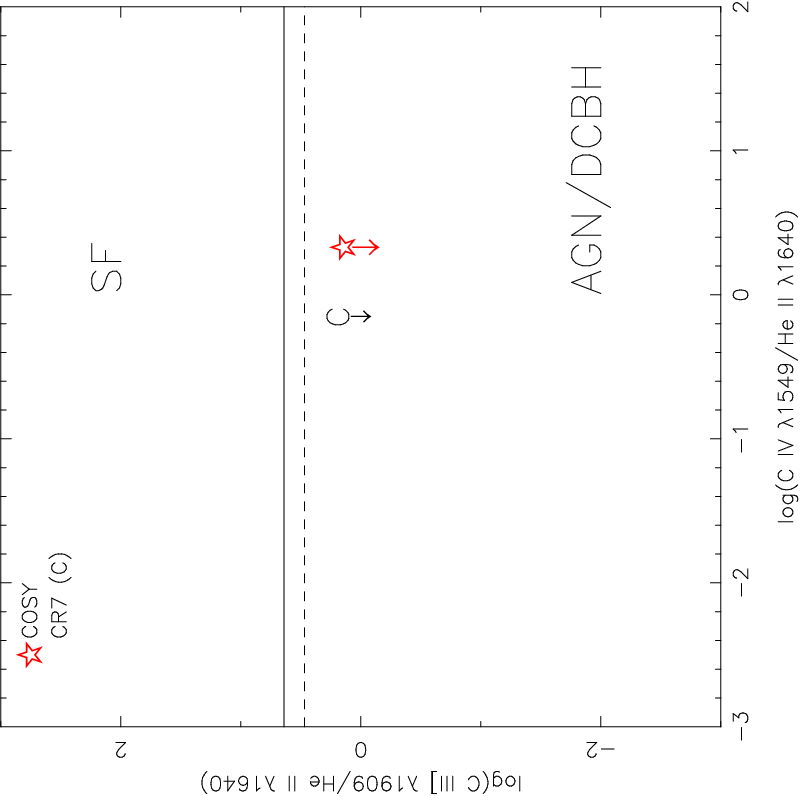}\vspace{0.05\columnwidth}
\includegraphics[angle=-90,width=0.42\columnwidth]{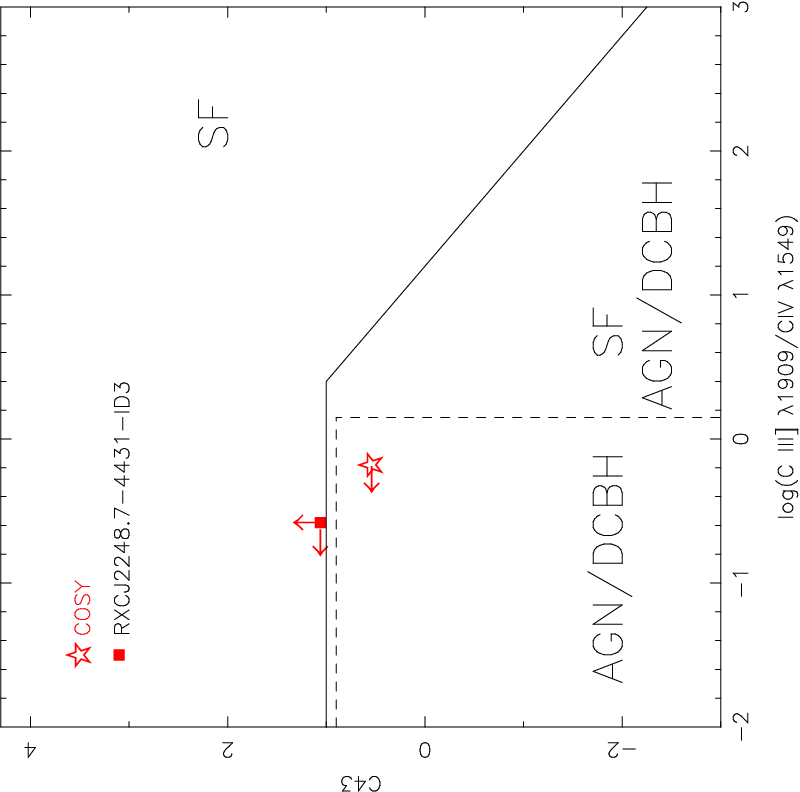}\hspace{0.05\columnwidth}
\includegraphics[angle=-90,width=0.42\columnwidth]{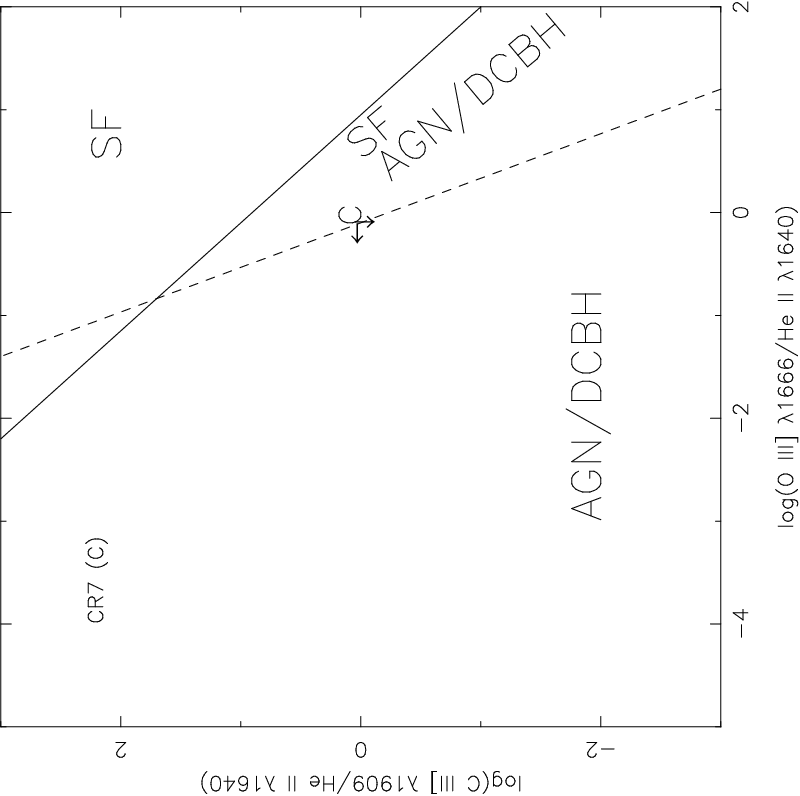}\vspace{0.05\columnwidth}
\includegraphics[angle=-90,width=0.42\columnwidth]{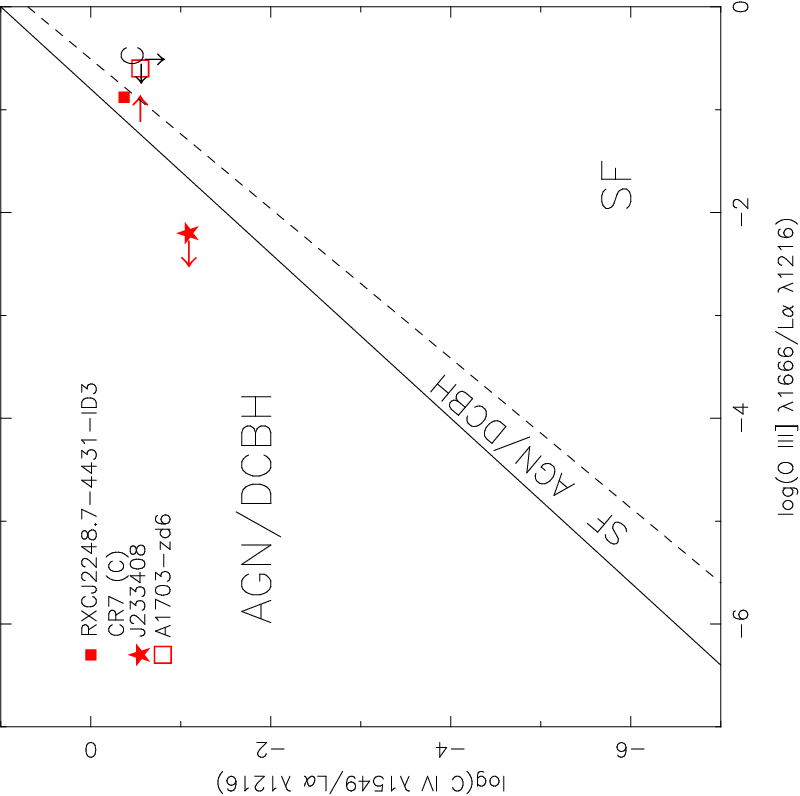}\hspace{0.05\columnwidth}
 \caption{Diagnostic diagrams considering the emission line ratios of the sample
 of LAEs (see Sect.~\ref{dlae}). Line curves separate SFs zone from that of AGNs as defined
 in Figs.~\ref{fr2} and \ref{fr4}.  Each LAE is represented by different symbol
 as indicated. Arrows indicate that only the upper and/or lower limit value of a given emission-line ratio was quoted.} 
\label{fr5}
\end{figure*}

Similar diagnostic diagrams as the ones in  Figs.\ \ref{fr2} and \ref{fr4} are plotted in Fig.\ \ref{fr5}
 for the LAE sample using the line fluxes compiled from the literature and listed in Table~\ref{tab1}. 
In order to classify the LAEs using these diagnostic diagrams, we included  zone-separation curves obtained using the control sample 
(empirical classification, see Fig.~\ref{fr2}) and those from the diagrams containing photoionization models (theoretical classification, see Fig.~\ref{fr4}). 
The \ion{C}{iv}/Ly$\alpha$ vs. \ion{N}{v}/Ly$\alpha$ diagnostic diagram was not included in this figure since it does not provide a method to distinguish AGN-like from SF-like objects.

Table~\ref{tab2} lists the classification for each LAE based on each diagnostic diagram, and the resulting classification, i.e. the frequency of the highest occurrence of a given object class. 
We marked in  Table~\ref{tab2}, between parentheses, the alternative classification that could be derived if we take into account the arrows (added when only upper or lower limits of the line ratios are available) and the intrinsic errors in the line-curves determinations.
Three of the LAEs in our sample (CR7\,(C), J233408 and COSY) can be classified as AGN-like objects using our diagnostic diagrams and {two of them, ID3 and A1703-zd6}, 
can be classified as SF-like objects.

\subsection{Detailed models}
\label{detal}

Once the spectral classification of the LAEs is adopted (see Table~\ref{tab2}),
it is possible to derive some parameters describing them by building detailed photoionization models.
The standard procedure is to reproduce  a  ratio between a given metal emission line
and one hydrogen emission line used as reference (e.g. \citealt{dors11, dors17a, contini17}). 
However, since  in some cases the Lyman continuum can  escape from a galaxy
 \citep[e.g.][]{rivera17}  and due to the  scattering of the Ly$\alpha$ photons by neutral hydrogen
  \citep[e.g.][]{hayes15},  we used the intensities of metal lines in relation to the \ion{He}{ii} $\lambda1640\rm\ \AA$ 
  as an observational constraint. 

 The  methodology  adopted  is similar to that of by \citet{dors17a}, who
built detailed photoionization models to reproduce optical emission lines of
Seyfert 2s in order to derive  their chemical  abundances. 
An  initial model assuming the following input parameters  
was built with:
\begin{enumerate}
\item Metallicity -- An initial value $Z/Z_{\odot}=0.1$ was assumed. This value is in the  metallicity range
derived, for example, for  CR7 \citep{sobral17, agarwal17} and  for  A1703-zd6 \citep{stark15}.
\item Number of ionizing photons -- From Fig.~\ref{fr1}, we found that only models with 
$\log Q(\rm H) \: \geq \: 53$  predict Ly$\alpha$ and \ion{He}{ii}
luminosity observed  for the objects in our sample of LAEs. Thus, we adopted  $\log Q(\rm H)=53$ as an initial value for the number of ionizing photons.
\item SED -- For each LAE,  the ionizing source type  was assumed to be the one derived in the  adopted classification
listed in Table~\ref{tab2}. For ID3 {and  A1703-zd6, the  objects in the sample classified as  SF},
we adopted the SED for a stellar cluster with an age of $10^{4}$ yr (see Section~\ref{spopii}). 
For the objects classified as  AGN-like, {we  adopted both DCBH  ($M=10^{6} \: M_{\odot};\dot{M}=1.0)$
and power law ($\alpha_{\rm ox}=-0.8$)  SEDs. } 
\item N and C abundances -- The N and C abundances were calculated from{ Eqs.~\ref{eq1a} and \ref{eq2}, respectively.}
\item Electron density -- The value of the electron density was considered to be $N_{\rm e}=500 \: \rm cm^{-3}$.
The electron density in LAEs at high redshift is unknown and the value above only represents gas in the low density limit, where collisional de-excitation does not affect the line  formation.
\end{enumerate}
We then  ran new models varying  $Z$, N/H, C/H values  separately in steps of $\pm$0.2 dex, which is the  typical uncertainty in  
nebular abundance estimations derived through  photoionization models \citep{dors11}. 
The $\log Q(\rm H)$ was varied considering a step of 0.5 dex. Only one parameter was varied at a time and the   
optimization method {\sc phymir} \citep{vanhoof97}  was employed to  select the best fitting  model for the set of emission line intensities. {For ID3, J233408 and A1703-zd6 the \ion{N}{v} line was not measured,
therefore,  the nitrogen abundance assumed in the models was scaled directly from the oxygen abundance 
by using  Equation~\ref{eq1a}.}

In Table~\ref{tab3}, the observed emission-line ratios  and the ones predicted by our detailed models built to represent each object of the sample are listed. The best fit parameters of the models
are listed in Table~\ref{tab4}. {The ionization parameter $U$, listed in Table~\ref{tab4} for each LAE, is not an 
input parameter of the  models. It is defined as $U= Q({\rm H})/4\pi R^{2}_{\rm in} n  c$, where $ Q({\rm H})$  
is the number of hydrogen ionizing photons emitted per second
by the ionizing source, $R_{\rm in}$  is  the distance from the ionization source to the inner surface
of the ionized gas cloud (in cm), $n$ is the  particle
density (in $\rm cm^{-3}$), and $c$ is the speed of light.}

{To estimate  the error in the parameters derived for each object, 
we adopted a similar methodology as the one employed by \citet{dors17a}.
 Firstly, we consider the final model solutions for each LAE (see Table~\ref{tab4}).
 From these models,  we vary the  metallicity (with a step of 0.05 dex), ionization parameter (step of 0.1 dex) and the nitrogen abundance (with a step of 0.05 dex) 
 in order to obtain the range of  parameter values for which the models reproduce the observational uncertainties of the emission line ratios
 of each LAE. This procedure was carried out only for COSY and CR7(C) because only these two objects have  measurements of the \ion{He}{ii}$\lambda1640$\AA\ 
 flux, making it possible to quote parameter values. However, carbon emission line fluxes
 were not measured in the spectrum of these objects, therefore, it was not possible to 
 estimate the uncertainty in the C abundance estimations.  
 For the other objects (ID3, J233408 and A1703-zd6), the  \ion{He}{ii}
 line flux was not constrained, therefore, the parameter values derived for them must be interpreted as upper limits.}

\begin{table*}
\caption{Comparison between the  logarithm of  observational emission-lines ratios (calculated using the fluxes listed in  Table~\ref{tab1}) 
with the ones predicted by our detailed photoionization models. The observed values compiled from the literature are  referred as "Obs." while the predicted values by the photoionization models as  "Mod."}
\vspace{0.3cm}
\label{tab3}
\tiny
\begin{tabular}{@{}l|ccc|cc|ccc|ccc|cc@{}}
\hline
                                                     & \multicolumn{3}{c|}{COSY}                                & \multicolumn{2}{c|}{ID3}                        & \multicolumn{3}{c|}{CR7(C)}                                             & \multicolumn{3}{c|}{J233408}                                                           &      \multicolumn{2}{c@{}}{A1703-zd6 }  \\
                                                      &    Obs.                      &     \multicolumn{2}{c|}{Mod.}       &         Obs.    &  \multicolumn{1}{c|}{Mod.}                                     &   Obs.                  & \multicolumn{2}{c|}{Mod.}                  &     Obs.         &    \multicolumn{2}{c|}{Mod.}                                        &    Obs.   & \multicolumn{1}{c@{}}{Mod.}  \\               
                                                   &
                                                                                                               &    AGN     & DCBH                  &                    &    SF                                          &                             & AGN             & DCBH                 &                                   &       AGN & DCBH                                   &                                        &  SF     \\
\hline
& &&& && &&& &&& &\\
$\rm \frac{Ly\alpha}{\ion{He}{ii}}$  &   $1.26\pm0.11$         &    1.29      &    1.75                  &      $>$1.34  &      2.88                                     &   0.41$\pm0.21$    &  1.19             &    1.70                 &                   $>$1.92    &     1.12    &   1.77                                   &  $>1.13$                          &  3.27  \\
                                                   &                                  &                &                             &                    &                                                  &                             &                      &                           &                                  &                &                                             &                                       &       \\
$\rm \frac{C\:III]}{He\:II}$             &    $<\:0.03$                &   $-$0.03  &    $-$0.35              &       ---         &      ---                                         &    $<-0.00$            &  $-0.69$         &  $-0.44$             &                    ---          &     ---        &   ---                                       &  ---                                   &  ---   \\ 
                                                  &                                   &               &                             &                    &                                                  &                             &                      &                           &                                  &                &                                              &                                             &        \\
$\rm \frac{N\:V}{He\:II}$              &  $0.30\pm0.12$           &   0.27      &   0.26                   &        ---        &     ---                                             &    $0.11\pm0.24$   &   0.14            &  $-0.01$          &                  ---            &    ---        &  ---                                           &   ---                                         & ---       \\
                                                 &                                    &               &                             &                    &                                                  &                             &                       &                           &                                  &               &                                              &                                                &         \\
$\rm \frac{C\:IV]}{He\:II}$           &   $< \: 0.33$                 &   0.30      &   0.30                    &       $>0.97$  &    0.97                                       &   $<-0.15$              &   $-0.19$         &  $-0.22$             &                   $>0.85$    &    0.83    &  0.88                                      &  $>0.57$                                &  0.57     \\
                                                 &                                    &                &                             &                      &                                                &                            &                       &                           &                                  &               &                                              &                                                &          \\
$\rm \frac{O\:III]}{He\:II}$          &   ---                               &    ---         &    ---                    &     $>0.46$     &    0.56                                      &   $<-0.09$           &   $-0.24$         &   $-0.12$               &                 ---             &  ---         &   ---                                      &    $>0.52$                               &  0.85     \\[4pt]                                          
\hline
\end{tabular}
\end{table*}

\begin{table*}
\caption{{Best fit nebular parameter used to reproduce} the emission line ratio observed in the LAE sample. Results for a same object  considering different ionizing sources  (AGN, DCBH, SF)  in the models are
presented separately. {The values of the parameters for ID3, J233408 and A1703-zd6 must be considered as upper limit values (see Sect.~\ref{detal}).}  }
\vspace{0.3cm}
\label{tab4}
\begin{tabular}{@{}l|rr|r|rr|rr|c@{}}
\hline              
                                       &    \multicolumn{2}{c|}{COSY}                                        &   \multicolumn{1}{c|}{ID3}                &    \multicolumn{2}{c|}{CR7(C)}                                    &      \multicolumn{2}{c|}{J233408}    &   \multicolumn{1}{c@{}}{A1703-zd6}  \\			   
Parameter$\backslash$SED                                        &    \multicolumn{1}{c}{AGN}                           &   \multicolumn{1}{c|}{DCBH}                             &       \multicolumn{1}{c|}{SF}              &  \multicolumn{1}{c}{AGN}                           &     \multicolumn{1}{c|}{DCBH}                            &   \multicolumn{1}{c}{AGN}          &    \multicolumn{1}{c|}{DCBH}                 &      \multicolumn{1}{c}{SF}   \\
\hline
	                 &                                      &                                         &                          &                                      &                                         &                    &                               &                   \\[-2pt]
$\log Q(\rm H)$                & $53.45^{+0.75}_{-0.30}$ &  $56.0_{-1.10}^{+0.50}$     &         55.98         &    $56.0_{-2.8}$               &     $55.8_{-2.0}$               &   55.9          &    56.0                     &        53.35  \\[4pt]
$\log U$                           & $-2.29^{+0.22}_{-0.19}$ &    $-1.36_{-0.16}^{+0.17}$   &        $-1.29$       &    $-2.03_{-0.53}$            &    $-1.40_{-0.65}$             &  $-1.53$      &   $-1.41$                 &      $-2.18$ \\[4pt]
12+log(O/H)                     &   $8.08_{-0.24}^{+0.51}$ &  $7.86^{+0.29}_{-0.59}$    &          8.36          &    $8.36^{+0.24}_{-0.97}$ &    $8.34^{+0.19}_{-0.65}$    &   8.36         &   7.83                      &       8.27     \\[4pt]
$Z/Z_{\odot}$                   &  $0.24_{-0.10}^{+0.56}$  &  $0.15^{+0.14}_{-0.11}$    &          0.46          &    $0.46^{+0.36}_{-0.41}$ &   $0.45^{+0.25}_{-0.35}$      &    0.47         &   0.14                      &        0.38     \\[4pt]                
$\rm \log(C/O)$                &    $<-0.40$                     &   $<-1.0$                         &       $-0.54$        &    $<-1.19$                      &     $<-1.12$                         &   $-0.18$     &   $-0.54$                 &      $-0.93$   \\[4pt]
$\rm \log(N/O)$                & $0.03^{+0.20}_{-0.08}$  &  $0.28^{+0.44}_{-0.24}$   &       $-1.39$         &    $-0.51^{+0.05}_{-0.63}$&     $0.14_{-0.44}^{+0.15}$    &    $-1.24$     &    $-1.24$                 &     $-1.39$    \\[4pt]
\hline
\end{tabular}
\end{table*}

\section{Discussion}
\label{disc}

\subsection{Ly$\alpha$ and \ion{He}{ii} emission}

The  Ly$\alpha$ and \ion{He}{ii} fluxes  play an important role in characterizing the 
SED of the embedded ionizing source of LAEs  (e.g. \citealt{sobral15}). 
In fact, the lack of \ion{He}{ii} emission in the majority of the LAE spectra indicates 
a drop in emission {at about 54.4 eV}, while  high Ly$\alpha$ luminosity  
requires  high $Q(\rm H)$ values. Therefore, the comparison between Ly$\alpha$ and \ion{He}{ii} fluxes 
in LAEs with those observed in objects with known nature or with photoionization model predictions,
 is the first step to uncover the nature of the LAE ionizing source. In the bottom panel of Fig.~\ref{fr1}, we 
see that the  AGN  observational data  (see Sect.~\ref{control})
form a clear linear trend, which can be represented by 
\begin{equation}
\log[L({\rm Ly\alpha})]=0.96 (\pm0.05) \: \log[L({\rm \ion{He}{ii}})]+2.56(\pm 2.23),
\end{equation}
with the high-$z$ radio galaxies occupying the upper end of this sequence.
It can be noted that the Ly$\alpha$ luminosity of the LAEs is  more consistent 
with the one of quasars and of less bright  high-$z$ radio galaxies, and they are brighter 
than    Seyfert 2  by a factor of  $\sim$100. 

Let us  define the ratio
\begin{equation}
\label{eq1}
R=\frac{L({\rm He\:II})}{L({\rm Ly\alpha})}
\end{equation}
as a feature of the field radiation. Concerning the    $R$ values for different classes 
of  AGNs from the control sample, for high-$z$ radio galaxies these are in the range 0.03-0.60, for Seyfert 2 in the range  
 0.01-0.3, and for Quasars in the range 0.06-0.30. 
From Table~\ref{tab1}, for CR7 C we derived $R=0.38$; a value consistent 
with the $R$ values for high-$z$ radio galaxies. For COSY we obtained $R=0.05$, 
consistent with the $R$ values of  all  classes of AGNs.
For other LAEs only  the upper limit for \ion{He}{ii} was defined, therefore 
the $R$ ratios were not calculated.
 
Now, we compare  $R$ values predicted by the models with the ones of the LAEs. The AGN and DCBH models predict   $R$ values in the range 0.002-0.11 and 0.006-0.012, respectively, while SF models estimate 0.0001-0.03. Thus, only AGN models   reproduce the $R$ value of COSY (0.05) and all models fail to reproduce that for  CR7 C (0.38).  The discrepancy  found for CR7 C  could be  due to Ly$\alpha$ escape from the nebulae, alternative Ly$\alpha$ production mechanisms, dust attenuation \citep{matthee17}, resonant scattering of  Ly$\alpha$,  and/or the presence of  PopIII stellar cluster as a secondary  ionizing source.

\subsection{LAE properties}
 
\subsubsection{ID3}

\citet{mainali17} found that  the ionizing source
of ID3 is probably a low metallicity stellar population. This result is mainly derived from  comparison
between the ID3 line ratio intensities \ion{He}{ii}/\ion{C}{iv} versus \ion{O}{iii}]/\ion{C}{iv};  
and \ion{C}{iv}/\ion{He}{ii} versus \ion{O}{iii}]/\ion{He}{ii} with observational 
data and photoionization predictions.  Based on the empirical and theoretical diagrams C43=log[(\ion{C}{iv}$\lambda$1549+\ion{C}{iii}]$\lambda$1909)/\ion{He}{ii}$\lambda$1640]
 versus \ion{C}{iii}]/\ion{C}{iv}; and
\ion{C}{iv}/Ly$\alpha$ versus \ion{O}{iii}]/Ly$\alpha$, we confirm
the result found by \citet{mainali17} where the main ionizing source of ID3
is a  young stellar cluster. Since the results above are derived using  an upper limit for \ion{He}{ii}$\lambda$1640,
these are somewhat uncertain and deeper spectroscopic data would be needed to unveil the
true nature of this object.

From the detailed modelling procedure
in Sect.~\ref{detal}, we derived  the number of ionizing photons  $\log Q(\rm H) \approx56$  which,
 taking into account the {\sc  STARBURST99} \citep{leitherer99} predictions for  instantaneous star formation, indicates the presence of ionizing 
 stellar cluster(s)  with  mass of $\sim 10^{9}$ $M_{\odot}$. The metallicity
derived for this object is ($Z/Z_{\odot})=0.46$, with $\rm \log(N/O)=-1.39$  and $\rm \log(C/O)=-0.54$.
\citet{mainali17} did not estimate these specific abundances for this object, but reported that ID3
has a low gas metallicity. In Fig.~\ref{f6d} (bottom panel), we compared  our ID3
predictions for  $\rm \log(C/O)$   vs. 12+log(O/H) with abundance estimates for local star-forming regions ($z<0.1$) and for
the star forming galaxy Q2343-BX418 ($z\sim2.3$),  calculated from direct measurements of the
electron temperature. We can see that  the estimates for ID3 are consistent with the  ones for  SFs.  Concerning the N/O  abundance ratio, in Fig.~\ref{f6d} (top panel), 
 our predictions are compared with those of local SFs as well as with estimates for Seyfert 2 AGNs, derived from photoionization models by \citet{dors17a}. Again, since it was not possible to fit the \ion{N}{v}/\ion{He}{ii} ratio for ID3, this object is not represented in the N/O-O/H diagram. The agreement  between element abundance estimates of ID3 and those seen for SFs additionally suggests that ID3 is a star forming galaxy.

\subsubsection{COSY}

\citet{laporte17} compared the observational emission line ratios \ion{C}{iii}]/\ion{He}{ii} versus \ion{N}{v}/\ion{He}{ii},
and versus \ion{N}{v}/\ion{C}{iv} of this object with those predicted by AGN and SF photoionization models. 
These authors found that COSY data is irreconcilable with
SF models and due to the high value of \ion{N}{v} emission line, the data is located
at the extreme end of the AGN model predictions. As can be seen in Figure~7 of
\citet{laporte17}, the photoionization models considered by these authors indicate that 
COSY has a logarithm of the ionization parameter higher than $-0.5$ and metallicity
$(Z/Z_{\odot})\: > \: 0.5$. From our analysis using all diagrams where it was possible to include COSY data, 
we obtain a non-thermal (AGN/DCBH) ionizing source for COSY in agreement with \citet{laporte17}. 

From the high value of the intensity of  the  \ion{N}{v}/\ion{He}{ii} ratio,  we derived a high N/O abundance
ratio for this object from our detailed modelling. In Fig.~\ref{f6d},  the log(C/O) and log(N/O) versus log(O/H) values obtained for 
COSY (see Table~\ref{tab4}) are compared with those obtained in SFs and AGNs.
Since only the upper limit of the carbon  lines  were quoted by \citet{laporte17}, it was only possible
to derive the upper limit for carbon abundance.  We can see in Fig.~\ref{f6d} that the C/O value obtained by 
the COSY AGN-model follows the tendency derived  for SFs, however, the value derived by  the DCBH-model
is out of the sequence. Unfortunately, carbon abundance determinations for AGNs are rare in the literature.
Concerning N/O, both AGN and DCBH model predictions are higher by $\sim 0.6$ dex
than those  of  SFs with similar oxygen abundances.  
The  N/O values  derived for COSY are higher than those derived for local Seyfert 2 nuclei  by \citet{dors17a}.
It is worth noting that  \citet{dors17a} pointed out that N/O and O/H abundances in Seyfert 2  are similar 
to those derived for local extragalactic disc \ion{H}{ii} regions with high metallicity. However, this result is not valid
when the N/O-O/H estimations for COSY are considered.

\subsubsection{CR7 clump C}

Concerning CR7 C, \citet{sobral17} compared  photoionization model predictions  with observed
line ratios and concluded that it may host a high ionization AGN, with low metallicity ($\sim \:0.05-0.2 \: Z_{\odot}$)
and high ionisation parameter ($\log U\approx-2.5$).
\citet{sobral17} pointed out that the results above were obtained if the barely measured \ion{N}{v}  line
is emitted by the clump C. \citet{matthee17} presented spectroscopic follow-up observations of CR7 with 
ALMA and,  based on [\ion{C}{ii}]158$\mu$m emission, inferred the same range of metallicity above. 
However, a very low metallicity ($\approx0.005 \: Z_{\odot}$)
was derived by \citet{bowler+17}. 

{Our analysis indicates that CR7 clump C hosts an AGN/DCBH, 
such as suggested by \cite{sobral17}.}
In Fig.~\ref{f6d} (bottom panel), it can be seen that our C/O abundance results for this object   is 
significantly lower than  the one derived for other objects with similar metallicity.
This is due to the relative low limit for the ratio \ion{C}{iv}/\ion{He}{ii}. 
The  N/O-O/H values derived from AGN-models are in consonance with those derived
for local Seyfert 2, however, the  DCBH-model predicts a higher N/O value. This is in fact also noted in COSY and 
 is due to the softer SED of DCBH as compared to the AGN (see  Fig.~\ref{f5}), which results in  lower \ion{N}{v} line intensities than AGN models with the same abundances. Thus,
 in the DCBH detailed models we must assume higher nitrogen abundance 
and ionization parameters (see Table~\ref{tab4}) than the ones considered
in the AGN models. 

\subsubsection{J233408}

\citet{shibuya18} found that the   constraints on  the line ratios  \ion{O}{iii}]$\lambda$1663/\ion{C}{iv}$\lambda$1549 versus
\ion{He}{ii}$\lambda$1640/\ion{C}{iv}$\lambda$1549 of J233408 are compatible  with the predictions of 
SF  models of  \citet{feltre16}.
Our classification, based on only the  diagram \ion{C}{iv}/Ly$\alpha$ versus \ion{O}{iii}]/Ly$\alpha$, indicates that
J233408 hosts an AGN/DCBH. However, due to the few number of objects in the control sample (see Fig.~\ref{fr2}) and the problem related to Ly$\alpha$ leakage/scattering,
our LAE classification could be rather uncertain.

The metallicity derived from the detailed modelling was $Z/Z_{\odot} = 0.46$~(0.14)  for 
AGN (DCBH) models, respectively. Until now, these are  the only metallicity estimates  available in literature for J233408.
The detailed  modelling was based on  the fitting of the \ion{C}{iv}/\ion{He}{ii} ratio, which makes the abundance results somewhat uncertain.
In Fig.~\ref{f6d} (bottom panel), it can be seen that the C/O abundance for J233408  derived 
in AGN and DCBH models follows the trends derived for the SFs. Since neither measurements nor limits 
 for the \ion{N}{v} lines are available, the N/O  value in the models was considered to be solar and we refrain from representing it in Fig.~\ref{f6d} (upper panel).

\subsubsection{A1703-zd6}

\begin{figure}
\centering
\includegraphics[angle=-90,width=1\columnwidth]{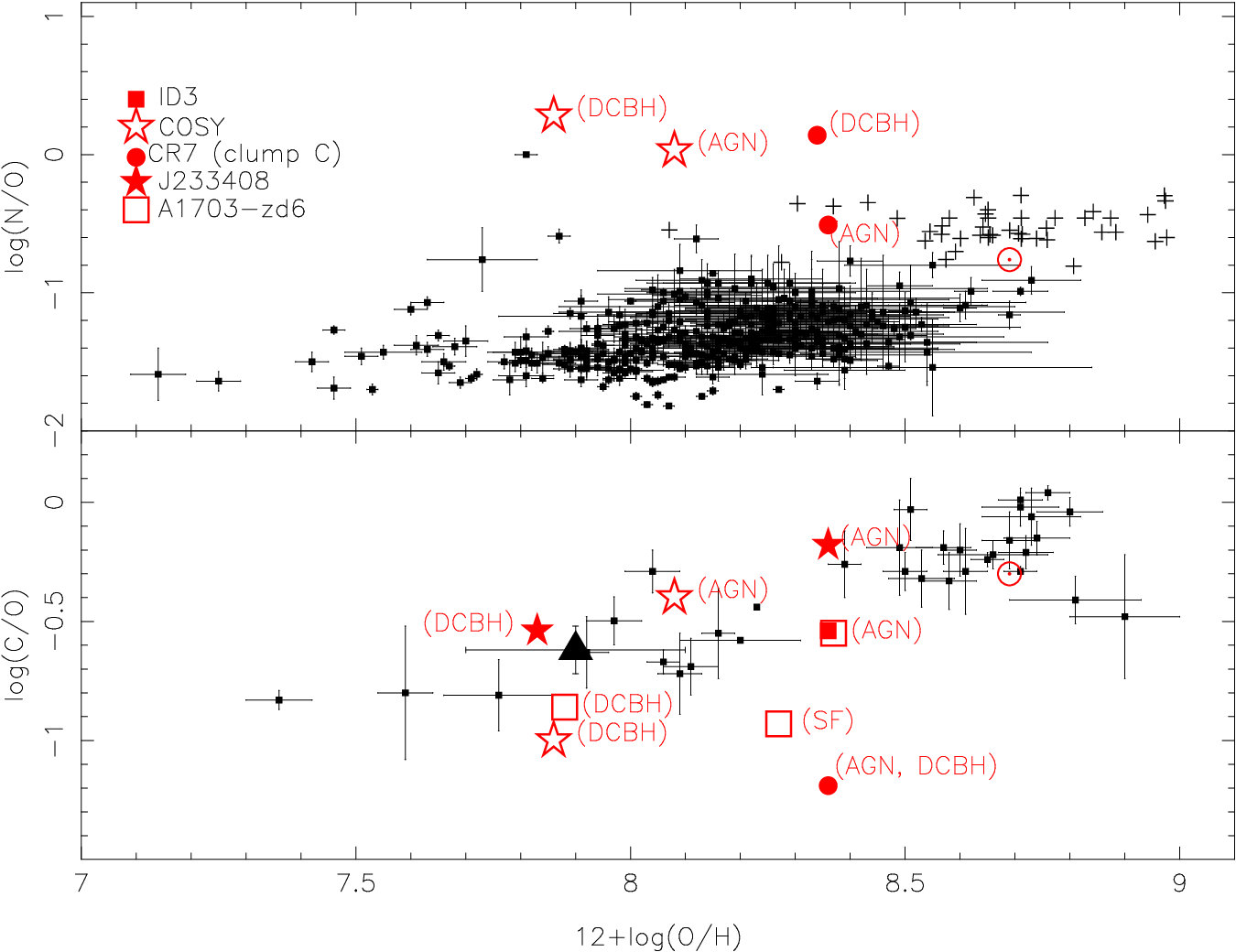}
\caption{Bottom:  log(C/O) versus 12+log(O/H). Abundance ratios  predicted by  the models  for the LAE sample, listed in Table~\ref{tab4},
are indicated in the plots.    The AGN, DCBH and/or  SF model
predictions were  obtained for a given LAE (see Table~\ref{tab4}), these are  indicated in the plot.
Squares represent estimations from optical-emission lines of star-forming regions by  
\citet{esteban02, esteban04, esteban05, esteban09}, \citet{kobulnicky98}  and \citet{garnett95,  garnett97}.  Triangle represents the
estimation for the galaxy Q2343-BX418 ($z\approx2.3$) by  \citet{erb10}. The   solar symbols (in red) represent the values for
solar abundance taken from \citet{allende-prieto01, allende-prieto02} and \citet{holweger01}. 
Top:  log(N/O) versus 12+log(O/H). Black symbols are  as in bottom panel but  data were  taken from \citet{hagele08, Hagele+11, Hagele+12}, \citet{perez11}, \citet{guseva11}, 
\citet{esteban09}, \citet{kobulnicky98}, and \citet{izotov06}.  Crosses are estimations
 for local Seyfert 2   AGNs ($z \: < \:0.1$) based on photoionization models performed by \citet{dors17a}. 
 Only the N/O and O/H predictions for the LAEs in which was possible estimate the value for the
 line ratio \ion{N}{v}/\ion{He}{ii} (i.e. COSY and CR7 (C)) are represented in this diagram. The error in our abundance estimations are about 0.2 dex
 (see Sect.~\ref{detal}).}
\label{f6d}
\end{figure}

\citet{stark15} demonstrated that the spectral properties of  A1703-zd6 can be reproduced
by photoionization models of a young ionizing source which is very hot and metal poor,
 as well as an AGN. In both  cases these authors found a very low gas metallicity, lower
than $Z/Z_{\odot}\approx0.02$, with $\log U=-1.35^{+0.24}_{-0.40}$ for the SF models. 
Considering the diagnostic diagrams in Sect.~\ref{res}, we also 
 found that  A1703-zd6 can be classified as SF. Our detailed models indicate a higher
{metallicity ($Z/Z_{\odot}\approx 0.4$)} for this object than the one derived by \citet{stark15},   
but similar $U$ value.
The discrepancy between the $Z$ values above is probably due to the C/O abundance 
assumed in the models rather than the use of different SEDs
(\citealt{stark15} did not explicitly state the C/O  value used). In Fig.~\ref{f6d}, we can see that the 
C/O versus O/H  predictions   indicate a lower C/O abundance than the one derived for objects with similar metallicity.  
Again, since \ion{N}{v} line intensity data is not available, the N/O  value in the models was considered 
and it is not represented in Fig.~\ref{f6d} (upper panel).

\section{Conclusions}
\label{conc}

We compared the available observational ultraviolet emission line intensities of five
LAEs  located  at high  redshift $(5.7 \: < \: z \:  < \:7.2)$ with predictions from photoionization models built with an ionizing source representative of an AGN, 
a Direct Collapse Black Hole (DCBH), and stellar clusters of  Population II stars.
From our analysis, we conclude the following.
\begin{enumerate}
\item Based on  diagnostic  diagrams,  we concluded that  CR7 (clump C), HSC\,J233408+004403 and  COSY probably have a non-thermal
ionizing source (AGN or DCBH). RXC\,J2248.7-4431-ID3 and {A1703-zd6}
seem to have  a stellar cluster as ionizing source. {It must be noted that due to the few number and the low significance of the detected emission lines the classification
above  could be rather uncertain.}
\item Detailed photoionization model fittings indicate a metallicity range  of $0.1 \: \la \: (Z/Z_{\odot}) \: \la \: 0.5$ for the LAEs considered.
\item The C/O estimations for the LAE sample, in general, are consistent with those for local star forming objects.
\item In  most cases,  an overabundance of N/O  was derived for LAEs in relation to AGNs and SFs with similar metallicities.
\end{enumerate}


\label{lastpage}

\end{document}